%% file: main_3_.tex
\documentclass[trackchanges,twocolumn]{aastex701}
\usepackage{breqn}
\usepackage{booktabs}
\usepackage{threeparttable}
\usepackage{mathrsfs}
\usepackage{caption}
\usepackage{gensymb}
\usepackage{comment}
\usepackage{longtable}
\usepackage{tabularx}

\newenvironment{aascases}
{\left\{\begin{array}{ll}}
{\end{array}\right.}

\newcommand{\curlc}{\mathscr{C}}

\newcommand{\shenzhangfactor}{\curlc = 
    \begin{aascases}
        \frac{(p + 2)(p - 1/3)}{p + 2/3}, & \nu_a < \nu_m, \\
        \sqrt{2}(p + 1)
        \frac{
        G\left(\frac{3p + 22}{12}\right)
        G\left(\frac{3p + 2}{12}\right)
        }{
        G\left(\frac{3p + 19}{12}\right)
        G\left(\frac{3p - 1}{12}\right)
        }, & \nu_m < \nu_a
    \end{aascases}
}

\begin{document}

\title{Radio Observations of the Unusual Tidal Disruption Event AT\,2022wtn: a Fast and Highly Energetic Outflow}

\include{authors}

\begin{abstract}

We present multi-epoch, multi-frequency radio observations of the tidal disruption event (TDE) AT\,2022wtn, obtained with the Karl G. Jansky Very Large Array (VLA) and Giant Metrewave Radio Telescope (GMRT), spanning 97-866 days after optical detection. The peak radio flux density increases until 300 days post optical discovery, flattens out for several hundred days, then begins to decrease at 534 days. Utilizing an updated equipartition analysis framework, we estimate several physical parameters of the event and the surrounding medium. We model AT\,2022wtn with two different geometries: a spherical and a conical emitting region. The spherical outflow model gives an expansion velocity of $v\approx0.21c$ and a kinetic energy of $\sim3.8\times10^{49}$ erg, and the conical outflow model yields a higher energy ($\sim1.8\times10^{50}$) and velocity ($v\approx0.41c$) than the spherical case. After ruling out the possibility of a relativistic jet, we consider several potential origins for sub-relativistic outflow regions in TDEs including unbound debris streams, collisionally-induced outflows, an accretion-driven wind, and an outflow from an accretion disk state transition, and find only an accretion disk state transition outflow to be consistent with the high energy and velocity found in our equipartition results. AT\,2022wtn is a uniquely powerful non-relativistic radio-emitting TDE, and joins a growing population that display a diverse range of outflow properties.

\end{abstract}


\section{Introduction} 

\begin{table*}[ht!]
\centering
\caption{Radio observations of AT\,2022wtn taken with the VLA and GMRT.}
\label{tab:observations}
\hspace*{-1.9cm}
\begin{threeparttable}
\begin{tabular}{cccccc}
\hline \hline
Epoch & Date (UTC) & $\delta t$ (d) & Telescope & Configuration & Frequency Bands \\
\hline 
1 & 2023-01-07 & 97 & VLA & C$\rightarrow$B & Ku \\
2 & 2023-03-20 & 169 & VLA & B & Ku \\
3 & 2023-04-17 & 197 & VLA & B & L, S, C, X, Ku, K \\ 
4 & 2023-05-23 & 233 & VLA & B & L, S, C, X, Ku, K \\
5 & 2023-07-29 & 300 & VLA & A & L, S, C, X, Ku, K \\
6 & 2023-09-09 & 342 & VLA & A & L, S, C, X, Ku, K \\
7 & 2024-03-19 & 534 & VLA & C & S, C, X, Ku \\
8 & 2024-08-26 & 694 & VLA & B & L, S, C, X, Ku \\
9 & 2024-12-09 & 799 & VLA & A & L, S, C, X, Ku \\
10 & 2025-01-20 & 841 & VLA & A & L, S, C, X, Ku \\
10 & 2025-02-14 & 866 & GMRT & -- & 4 \\
\hline
\noalign{\vskip 5pt}
\end{tabular}
\centering
\begin{tablenotes}[para]
\small
\centering
\parbox{0.8\textwidth}{%
Note: $\delta t$ is measured with respect to the estimated outflow launch date in the observer frame, $t_0=$ MJD 59854. Flux density measurements are given in Table \ref{tab:obsA}.
}
\end{tablenotes}
\end{threeparttable}
\end{table*}

On occasion, a star will venture too close to a supermassive black hole (SMBH) and be ripped apart by tidal forces overcoming the star's self-gravity \citep{rees_tidal_1988,hills_possible_1975}. In the aftermath, a luminous transient is produced that can be observed across the electromagnetic spectrum; such an occurrence is known as a tidal disruption event (TDE). During this process, roughly half of the resulting stellar debris is accreted onto the black hole, while the other half escapes the gravitational pull of the SMBH and is ejected into the surrounding environment \citep{1989Evans,guillochon_hydrodynamical_2013}. Most TDEs are observed in locations coincident with the nucleus of their host galaxies (\citealp{van_Velzen_2020}; but see also, \citealp{2018NatAs...2..656L,sfaradi_first_2025,yao2025massiveblackhole08,2025arXiv250109580J, 2025arXiv251202147L,2025NatAs...9..702Z}).

TDEs provide an excellent way to study accretion disk formation, relativistic jets, and outflow mechanisms around black holes. A variety of outflow mechanisms have been proposed in TDEs to explain radio observations, which constrain physical properties such as the outflow velocity, blast wave radius, and the density of the surrounding environment \citep{chevalier_synchrotron_1998,Metzger_2012,margalit_thermal_2021,matsumoto_generalized_2023,De_Colle_2012}. 

A small portion ($\lesssim1\%$) of the known TDE population have launched prompt on-axis relativistic jets \citep[e.g.,][]{zauderer_birth_2011,bradley_cenko_swift_2012,eftekhari_radio_2018,2023_2022cmc,De_Colle_2012} that generate extremely luminous ($\nu L_{\nu} \gtrsim 10^{40}$ erg/s) radio emission still detectable several years after initial disruption. Most ($\gtrsim99\%$) radio-detected TDEs exhibit less-luminous outflows ($\nu L_{\nu} \lesssim 10^{40}$ erg/s), commonly attributed to a quasi-spherical Newtonian outflow interacting with the circumnuclear environment (e.g., \citealp{alexander_discovery_2016,christy_dichotomy_2025,cendes_radio_2021}). These outflows have been well described by accretion-driven wind, collisions in the debris stream, the unbound debris interacting with the surrounding medium, or state transitions in the accretion disk launching an outflow \citep{alexander_discovery_2016,krolik_asassn-14li_2016,Bonnerot_2020,wu2025delayedradioemissiontidal,dai_unified_2018}, but the relative importance of these outflow mechanisms in powering the observed radio emission remains uncertain. By deriving the inferred physical properties of the event, we can constrain the outflow mechanism in the context of these models, though several (e.g., \citealp{goodwin_at2019azh_2022}) or none (e.g., \citealp{christy_peculiar_2024}) of them may sufficiently explain the outflow properties. As many as $40\%$ of radio-TDEs only display radio emission at late times, hundreds to thousands of days after the initial disruption \citep{alexander_radio_2020,cendes_ubiquitous_2024}.

In this paper we present multi-epoch, multi-frequency radio observations of the TDE AT\,2022wtn extending over 2 years after discovery. \citet{onori_case_2025} previously studied the optical, UV, and X-ray emission and found evidence for a prompt accretion disk formation and an edge-on viewing angle. We consider outflow models that can self-consistently explain the radio emission in the context of the available multiwavelength data for this event. 

The paper is organized as follows. In \S\ref{sec:observations} we describe our radio data and the reduction methods used. In \S\ref{sec:emodeling} we model the radio spectral energy distribution as synchrotron radiation. In \S\ref{sec:omodeling}, we carry out an equipartition analysis to derive physical properties of the outflow and the surrounding environment. In \S\ref{sec:discussion} we discuss implications of our modeling and look at AT\,2022wtn in the context of multiwavelength observations. In \S\ref{sec:conclusion} we summarize our results and present conclusions. All times and frequencies reported are in the observer frame, unless otherwise noted.

\begin{figure*}[ht!]
\centering
\includegraphics[width=\linewidth]{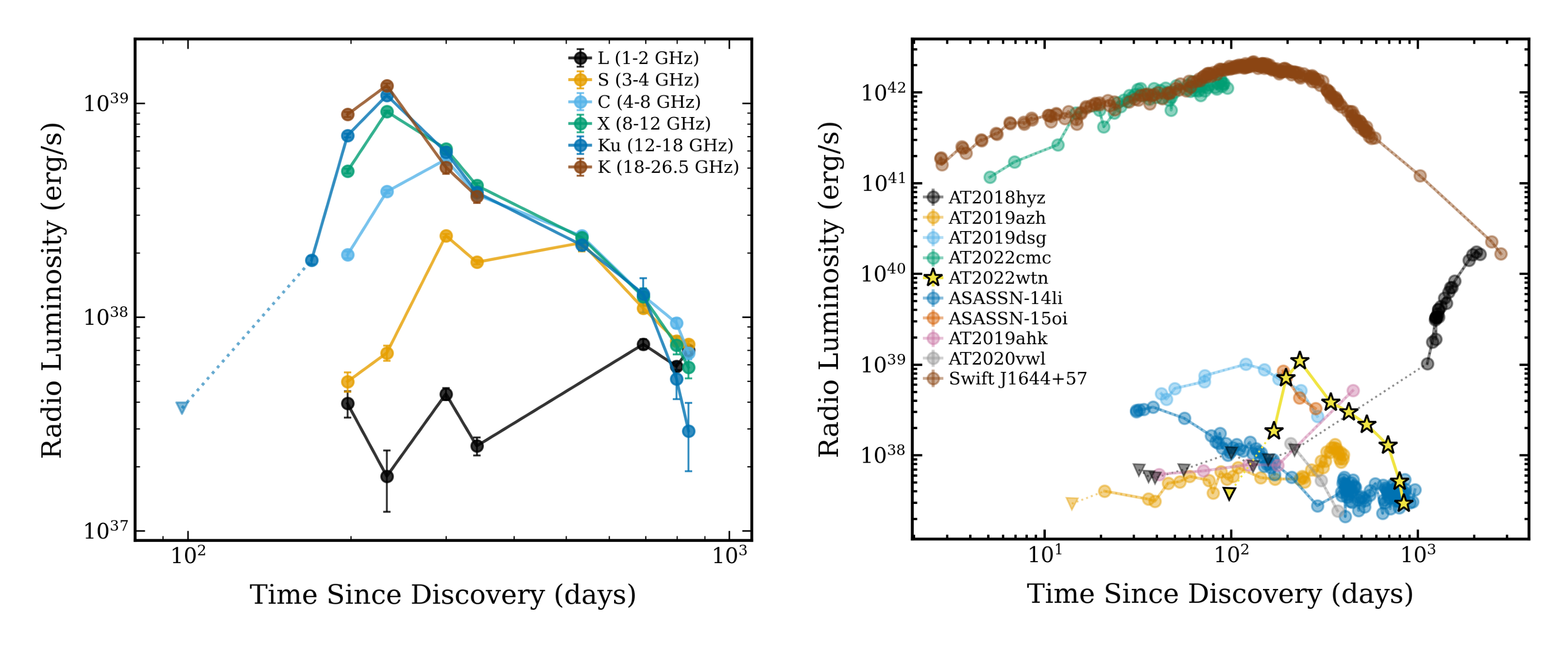}
\caption{
Left: radio luminosity light curves of AT\,2022wtn in several frequency bands. The early triangular data point in the Ku-band is a $3\sigma$ upper limit. The flux increases on different timescales in each band, but there is a clear brightening of the source at $\delta t\sim 200$ days. Right: radio luminosity vs. rest frame time of AT\,2022wtn in Ku-band (12-18 GHz) compared to a sample population of TDEs with radio emission in the same band. The sample includes ASASSN-14li \citep{alexander_discovery_2016}, AT\,2018hyz \citep{cendes_continued_2025}, AT\,2019azh \citep{goodwin_at2019azh_2022}, AT\,2019dsg \citep{cendes_ubiquitous_2024}, ASASSN-15oi \citep{hajela_eight_2025}, AT\,2019ahk \citep{christy_peculiar_2024}, AT\,2020vwl \citep{goodwin_second_2025}, Swift J1644+57 \citep{zauderer_birth_2011,eftekhari_radio_2018}, and AT\,2022cmc \citep{2023_2022cmc}. This plot was generated with the OTTER API \citep{franz_python_2026, franz_open_2026}}.
\label{fig:fig1and2}
\end{figure*}

\section{Observations} \label{sec:observations}

AT\,2022wtn ($z=0.049$; TNS\footnote{https://www.wis-tns.org/} Classification Report No. 13873, \citealp{2022TNSCR3389....1F}) was first discovered by the Zwicky Transient Facility \citep[ZTF;][]{bellm_zwicky_2018} on 2022 October 2 (all times and dates are given in UT). It was subsequently classified as a TDE on 2022 November 21 based on optical spectroscopic observations \citep{2022TNSCR3389....1F}. The transient's radio coordinates (RA (J2000) = 23:23:23.79, DEC (J2000) = +10:41:07.75 with an uncertainty of 0.04 arcseconds in each coordinate; \citealp{2023ATel15972....1C}) are consistent with the center of the galaxy SDSS J232323.79+104107.7, the less massive galaxy in a pair currently experiencing a merger\footnote{Merging galaxies may display an enhanced TDE rate due to significant changes in galactic structure and dynamics \citep{onori_case_2025}.} \citep{onori_case_2025}.

For our analysis, we define $\delta t$ to be the time since the optical discovery in the observer frame, MJD 59854.26 (TNS Astronomical Transient Report No. 160738 by ZTF; \citealp{2022TNSCR3389....1F}). After the initial classification, we began a 2+ year radio monitoring campaign, starting with an initial non-detection at $\delta t = 97$ days at a mean frequency of 15 GHz, followed by a significant brightening at $\delta t = 167$ days in the same band, confirming the transient nature of the emission \citep{2023ATel15972....1C}. We then began to observe AT\,2022wtn at multiple frequencies to constrain the broadband spectrum, starting at $\delta t = 197$ days (Table \ref{tab:observations}). We assume the luminosity distance to be $d_L=225.1$ Mpc based on a flat $\Lambda$CDM cosmology with $H_0=67.66$ km s$^{-1}$ Mpc $^{-1}$, $\Omega_m=0.31$, and $\Omega_\Lambda=0.69$ \citep{Planck18}.

\begin{figure*}[ht!]
\centering
\includegraphics[width=\linewidth]{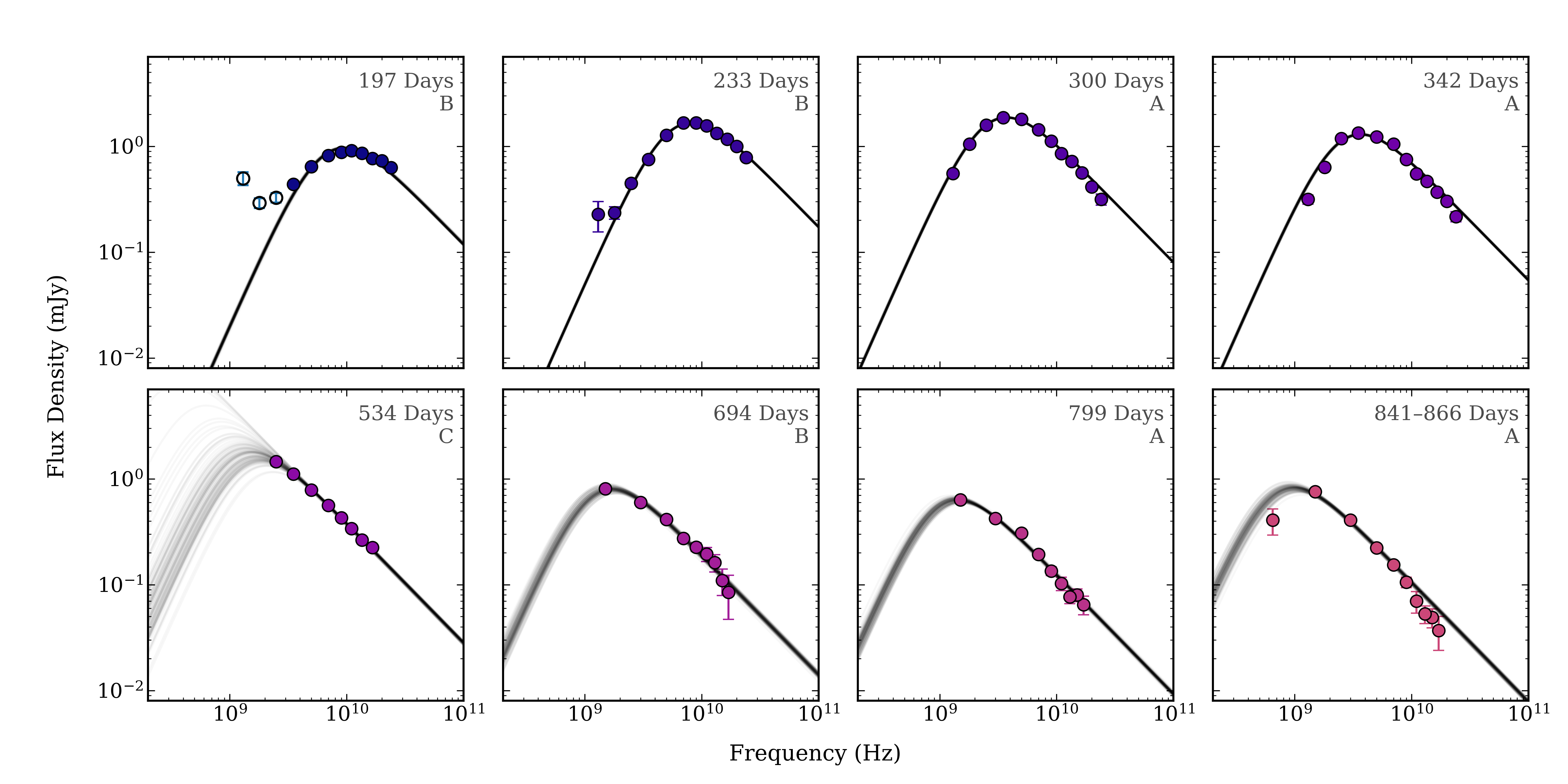}
\caption{
Radio SEDs from our VLA and GMRT data over eight epochs. The GMRT data point is added to the eighth SED fit due to the observations being close in time ($\delta t = 841$ and $\delta t = 866$), as indicated in Table \ref{tab:observations}. The lines in each plot are 100 sampled MCMC SED fits. Each plot shows the date in the upper-right corner, along with the VLA configuration that the epoch was taken in. The open circles in the first SED, corresponding to an excess transient component at low frequencies, indicate detections that were excluded from the SED fits (Section \ref{subsec:seds}).}
\label{fig:mcmc}
\end{figure*}

\subsection{VLA Data} \label{subsec:vla_data}

We obtained eleven epochs of data of AT\,2022wtn with the NSF's Karl G. Jansky Very Large Array (VLA), spanning from 2023 Jan 7 to 2025 Jan 20 ($\delta t = 97-841$ days). Data were obtained across multiple observing bands: L = 1-2 GHz, S = 2-4 GHz, C = 4-8 GHz, X = 8-12 GHz, Ku = 12-18 GHz, and K = 18-26.5 GHz under the VLA programs 20B-377 (PI: K. Alexander) and 24B-108 (PI: A. Goodwin). One epoch taken in the D-configuration is not discussed further: at lower frequencies, the transient was blended with a radio source $\sim8''$ away located in the nucleus of the interacting galaxy SDSS J232323.37+104101.7. 

We reduced the data in the Common Astronomy Software Application package (CASA, 6.5.5-21; \citealp{2007Casa,2022Casa}) with the VLA calibration pipeline (version 2023.1.0.124). 3C48 was used as the flux density calibrator and J2330+1100 as complex gain calibrator in all frequency bands and epochs. We also imaged the data in CASA, using the task \verb|tclean|. Frequency bands were split into 2 GHz sub-bands when the target was sufficiently bright. To obtain flux densities and uncertainties, an elliptical gaussian fixed to the size of the synthesized beam was fitted to the source using the task \verb|imfit|. The associated errors are a 1$\sigma$ statistical uncertainty as well as a 5 percent systematic uncertainty, caused by flux-density bootstrapping done during the calibration process \citep{Perley_Butler_2017}. The flux densities obtained, along with these uncertainties, are reported in Appendix \ref{app:fluxdensity}. 

\subsection{GMRT Data} \label{subsec:gmrt_data}

We also obtained two epochs of observations of AT\,2022wtn from the Giant Metrewave Radio Telescope (GMRT), on 21 Oct 2024 and 14 Feb 2025 under the program 47\_051 (PI: A. Goodwin), in the observing bands 4 (550-850 MHz) and 5 (1-1.45 GHz). Inspection of the first epoch revealed large and rapid phase variations on time scales shorter than the calibrator-target cycle times at levels that were impossible to remove using self-calibration due to insufficient signal-to-noise in images of the target field, rendering these observations unusable. The second dataset, which was successfully reduced, was taken in band 4 (550-850 MHz) only. 3C147 was employed as the flux density calibrator, and J2330+1100 as the phase calibrator. The data were manually reduced in CASA using standard flagging and calibration procedures. Self-calibration was used to correct for residual phase errors. Like the VLA data, an elliptical gaussian fixed to the size of the synthesized beam was fitted to the source using the CASA task \verb|imfit|. The results are presented in Appendix \ref{app:fluxdensity}, with the measured uncertainty including the
statistical uncertainty as well as a systematic uncertainty from flux-density bootstrapping (estimated at 5 percent). The results of the observations are summarized in Appendix \ref{app:fluxdensity} and shown in Figure \ref{fig:fig1and2}. 

\section{Emission Modeling} \label{sec:emodeling}

\subsection{Light Curve} \label{subsec:lightcurve}

AT\,2022wtn is a low-luminosity ($\nu L_{\nu} \lesssim 10^{40}$ erg/s) radio TDE, comparable to a majority of radio-TDEs \citep{alexander_radio_2020}. This is in contrast with the ultra-luminous jetted TDEs Swift J1644+57 \citep{eftekhari_radio_2018,zauderer_birth_2011} and AT\,2022cmc \citep{2023_2022cmc}, and the candidate off-axis jetted TDE AT\,2018hyz \citep{cendes_continued_2025}. The Ku-band radio light curve of AT\,2022wtn exhibits a fast initial rise after the first detection at $\delta t=167$ days, followed by a monotonic decline for the remainder of the observations (Figure \ref{fig:fig1and2}). The Ku-band luminosity rises by a factor of $\sim4$ between the first and second detections.
Due to the continued rise in peak flux from 197 days to 300 days (Figure~\ref{fig:fig1and2}), the optically thin SED above the peak at 197 days (Figure~\ref{fig:mcmc}), and an absence of steep spectra below the peak (Appendix~\ref{app:freefree}), we consider an origin in free-free absorption an unlikely explanation for the rising light curve.
Coupled with the non-detection at $\delta t=97$ days, this rapid rise is instead suggestive of an outflow launched several months post-disruption (for more discussion on the launch date, see Section \ref{sec:omodeling}).

\begin{figure}[t!]
\centering
\includegraphics[width=\linewidth]{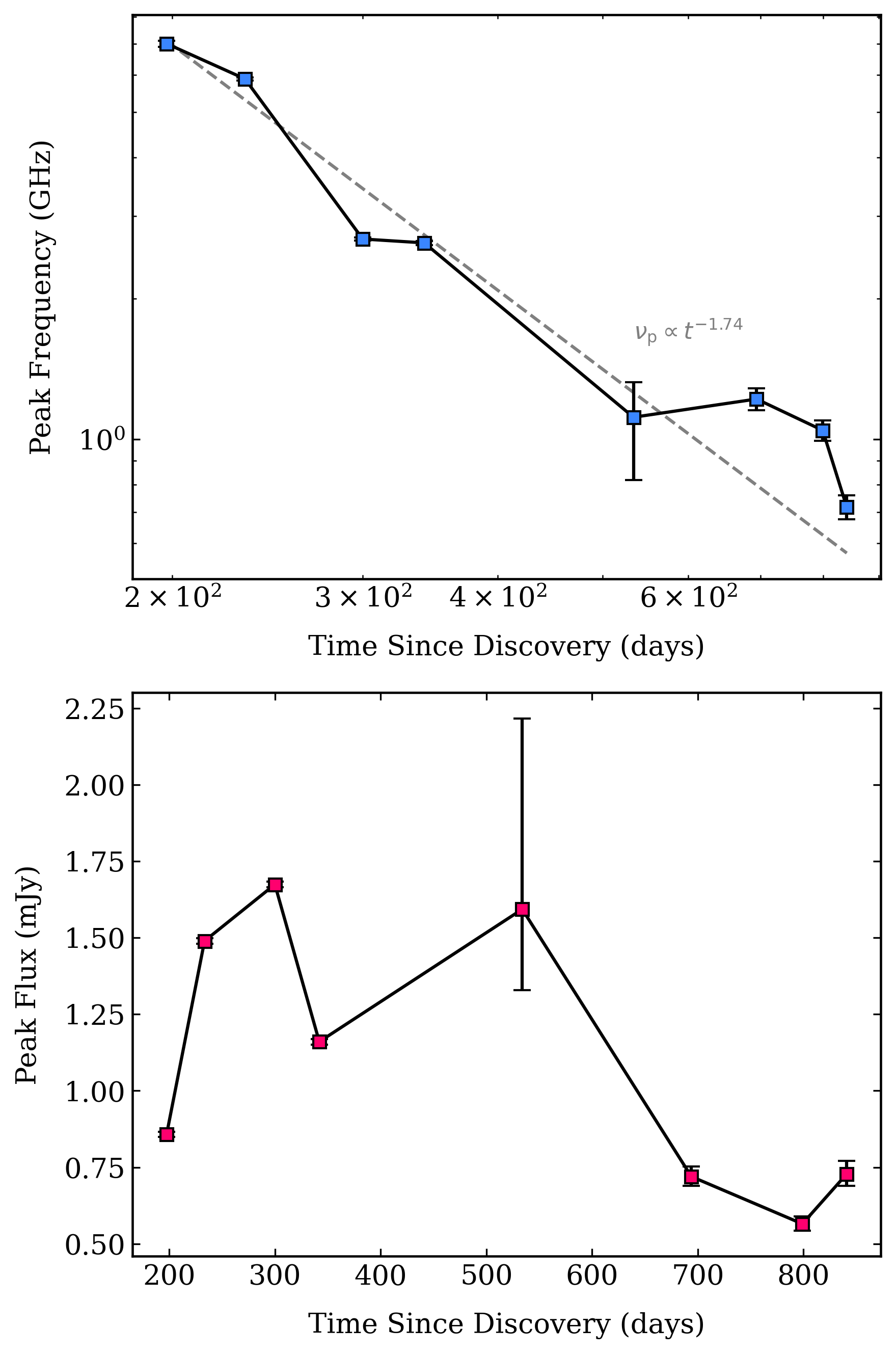}
\caption{
Peak frequency ($\nu_p$, top) and peak flux ($F_p$, bottom) through time}. We assume $\nu_p=\nu_a$ for our analysis. The values shown in these plots are obtained by setting $p$ to the average value of $p\approx3.26$. While the peak frequency decreases through time, the peak flux initially rises, then begins to decline.
\label{fig:peakvalues}
\end{figure}

\subsection{SEDs} \label{subsec:seds}

Following past works \citep[e.g.,][]{christy_peculiar_2024,goodwin_second_2025,alexander_discovery_2016,zauderer_birth_2011}, we model the radio spectral energy distributions (SEDs) of each epoch as a self-absorbed synchrotron spectrum, following the model presented in \citet{granot_shape_2002} (see Figure \ref{fig:mcmc}). For AT\,2022wtn, we assume that the break frequencies follow the ordering $\nu_m < \nu_a < \nu_c$, where $\nu_m$ is the synchrotron minimum frequency, $\nu_a$ is the self-absorption frequency, and $\nu_c$ is the cooling frequency. This is in line with other low-luminosity radio TDEs in the literature (e.g., \citealp{alexander_discovery_2016,goodwin_at2019azh_2022}), and given that the SEDs exhibit sharp peaks in each epoch, this is consistent with $\nu_a$ (but not $\nu_{\rm m}$) being observed in our frequency range. One can verify this by calculating the position of $\nu_{\rm m}$ (and $\nu_{\rm c}$) independently; see Appendix \ref{app:selfcons}. We model the flux density with a single broken power-law: 

\begin{dmath}
    F_\nu = F_{\nu\mathrm{,ext}} \left[\left(\frac{\nu}{\nu_a}\right)^{-s\beta_1} + \left(\frac{\nu}{\nu_a}\right)^{-s\beta_2}\right]^{-1/s} .
    \label{eq:eq1}
\end{dmath}

Here, $\nu$ is the frequency, $F_{\nu\mathrm{,ext}}$ is the normalization parameter, $\beta_1 = \frac{5}{2}$, and $\beta_2 = \frac{1-p}{2}$ ($p$ is the power-law energy index of shock-accelerated non-thermal electrons). Following the prescription in \citet{granot_shape_2002}, we employ $s = 1.25-0.18p$ for the smoothness of this break.

We fit the spectra for each epoch individually using \verb|emcee| \citep{emcee_main}, a Markov chain Monte Carlo (MCMC) Python module. To determine the model parameters, we used a Gaussian likelihood function and uniform priors. The three estimated parameters are $p$, $F_{\nu\mathrm{,ext}}$, and $\nu_a$; the employed priors were $1.5 < p < 4$, $-4 < \log(F_{\nu\mathrm{,ext}}/{\rm mJy}) < 2$, and $6 < \log(\nu_a/{\rm Hz}) < 11$ (e.g., \citealp{christy_peculiar_2024}). 

\begin{table}[t!]
\centering
\caption{SED Parameters}
\label{tab:sedparam}
\hspace*{-1.5cm}
\begin{tabular}{c c c}
$t$ (days) & $\nu_{\rm p}$ (GHz) & $F_{\rm p}$ (mJy) \\ \hline
197.6 & 6.85$^{+0.10}_{-0.10}$ & 0.86$^{+0.01}_{-0.01}$ \\
233.4 & 5.89$^{+0.05}_{-0.05}$ & 1.49$^{+0.01}_{-0.01}$ \\
300.0 & 2.68$^{+0.02}_{-0.02}$ & 1.67$^{+0.01}_{-0.01}$ \\
342.0 & 2.63$^{+0.03}_{-0.03}$ & 1.16$^{+0.01}_{-0.01}$ \\
534.0 & 1.12$^{+0.21}_{-0.29}$ & 1.59$^{+0.62}_{-0.26}$ \\
694.0 & 1.22$^{+0.07}_{-0.07}$ & 0.72$^{+0.03}_{-0.03}$ \\
799.0 & 1.05$^{+0.05}_{-0.05}$ & 0.56$^{+0.02}_{-0.02}$ \\
841.0 & 0.72$^{+0.04}_{-0.04}$ & 0.73$^{+0.04}_{-0.04}$ \\
\hline
\end{tabular}
\end{table}

\begin{table*}[ht!]
\centering
\caption{Derived physical parameters from our equipartition analysis for a spherical and conical geometry.}
\label{tab:equipartition}
\begin{tabular}{lccccccc}
\hline
Mode & $t$ (days) & $\log R_{\rm eq}$ (cm) & $\log E_{\rm eq}$ (erg) & $\log B$ (G) 
    & $\log N_e$ & $\log n_{\rm ext}$ (cm$^{-3}$) & $\beta$ \\
\hline
Spherical  & 197.6 & 16.59 $\pm$ 0.09 & 48.77 $\pm$ 0.60 & -0.01 $\pm$ 0.15 & 53.55 $\pm$ 0.60 & 3.15 $\pm$ 0.26 & 0.208 $\pm$ 0.033 \\
($f_A=1$) & 233.4 & 16.77 $\pm$ 0.09 & 49.10 $\pm$ 0.66 & -0.11 $\pm$ 0.16 & 53.87 $\pm$ 0.66 & 2.95 $\pm$ 0.29 & 0.197 $\pm$ 0.034 \\
 & 300.0 & 17.13 $\pm$ 0.10 & 49.51 $\pm$ 0.72 & -0.45 $\pm$ 0.17 & 54.29 $\pm$ 0.71 & 2.26 $\pm$ 0.31 & 0.253 $\pm$ 0.043 \\
 & 342.0 & 17.07 $\pm$ 0.09 & 49.32 $\pm$ 0.62 & -0.45 $\pm$ 0.15 & 54.10 $\pm$ 0.62 & 2.28 $\pm$ 0.27 & 0.187 $\pm$ 0.032 \\
 & 534.0 & 17.52 $\pm$ 0.23 & 49.94 $\pm$ 0.96 & -0.86 $\pm$ 0.21 & 54.71 $\pm$ 0.96 & 1.46 $\pm$ 0.41 & 0.252 $\pm$ 0.099 \\
 & 694.0 & 17.30 $\pm$ 0.10 & 49.38 $\pm$ 0.72 & -0.76 $\pm$ 0.16 & 54.16 $\pm$ 0.71 & 1.65 $\pm$ 0.29 & 0.126 $\pm$ 0.026 \\
 & 799.0 & 17.32 $\pm$ 0.09 & 49.35 $\pm$ 0.65 & -0.81 $\pm$ 0.15 & 54.13 $\pm$ 0.65 & 1.55 $\pm$ 0.28 & 0.113 $\pm$ 0.022 \\
 & 841.0 & 17.53 $\pm$ 0.10 & 49.61 $\pm$ 0.75 & -1.00 $\pm$ 0.17 & 54.39 $\pm$ 0.74 & 1.18 $\pm$ 0.31 & 0.165 $\pm$ 0.032 \\
\hline
Conical  & 197.6 & 17.02 $\pm$ 0.10 & 49.42 $\pm$ 0.74 & -0.32 $\pm$ 0.17 & 54.19 $\pm$ 0.73 & 4.13 $\pm$ 0.32 & 0.409 $\pm$ 0.053 \\
($f_A=0.1$) & 233.4 & 17.20 $\pm$ 0.10 & 49.78 $\pm$ 0.82 & -0.41 $\pm$ 0.18 & 54.56 $\pm$ 0.82 & 3.95 $\pm$ 0.34 & 0.397 $\pm$ 0.055 \\
 & 300.0 & 17.55 $\pm$ 0.10 & 50.13 $\pm$ 0.81 & -0.77 $\pm$ 0.18 & 54.91 $\pm$ 0.80 & 3.23 $\pm$ 0.34 & 0.470 $\pm$ 0.056 \\
 & 342.0 & 17.49 $\pm$ 0.10 & 49.96 $\pm$ 0.76 & -0.76 $\pm$ 0.18 & 54.73 $\pm$ 0.75 & 3.25 $\pm$ 0.33 & 0.378 $\pm$ 0.053 \\
 & 534.0 & 17.94 $\pm$ 0.23 & 50.56 $\pm$ 1.10 & -1.18 $\pm$ 0.22 & 55.34 $\pm$ 1.10 & 2.42 $\pm$ 0.43 & 0.469 $\pm$ 0.130 \\
 & 694.0 & 17.72 $\pm$ 0.10 & 50.07 $\pm$ 0.79 & -1.06 $\pm$ 0.18 & 54.85 $\pm$ 0.78 & 2.65 $\pm$ 0.34 & 0.278 $\pm$ 0.048 \\
 & 799.0 & 17.75 $\pm$ 0.10 & 50.02 $\pm$ 0.74 & -1.11 $\pm$ 0.18 & 54.79 $\pm$ 0.74 & 2.54 $\pm$ 0.33 & 0.254 $\pm$ 0.044 \\
 & 841.0 & 17.96 $\pm$ 0.10 & 50.29 $\pm$ 0.76 & -1.30 $\pm$ 0.18 & 55.06 $\pm$ 0.76 & 2.17 $\pm$ 0.33 & 0.342 $\pm$ 0.054 \\
\hline
\end{tabular}
\end{table*}

The posterior distributions for $F_{\nu\mathrm{,ext}}$, $\nu_a$, and $p$ were sampled with 100 walkers over 10,000 iterations, discarding the first 1000 steps to account for burn-in. The autocorrelation lengths are on the order of $\sim 30$ steps, with an acceptance fraction of $\approx0.7$, consistent with the recommended efficiency rates for ensemble samplers \citep{emcee_ascl}. We find some variation in $p$ with time when left as a free parameter, the most notable case being the first epoch at $\delta t =$ 197 days (see Appendix \ref{app:evolvingp}). Fluctuations in the synchrotron energy index are not expected in a single-zone synchrotron emission model. Furthermore, the fitted value of $p$ is highly dependent on the quality of the data (e.g., unrecoverable phase decorrelation at high frequencies) and is degenerate with the smoothing parameter, s. We, therefore, 
consider it unlikely that this variation is physical (Appendix \ref{app:evolvingp}) and adopt an average value of $p=3.26$ for our modeling and subsequent analysis, only fitting for $F_{\nu\mathrm{,ext}}$ and $\nu_a$ (see Table \ref{tab:sedparam}). We exclude the three lowest-frequency points from the SED fits in the $\delta t = 197$ days epoch that correspond to an excess transient component (see Section \ref{subsec:otheroutflow} for more discussion).

The peak frequency and peak flux density vs. time are shown in Figure \ref{fig:peakvalues}. The peak frequency decreases through time, roughly following a slope of $\nu_p\propto t^{-1.74}$. The peak flux density increases by a factor of 2 in between $\delta t =$ 197-300 days (see Figure \ref{fig:mcmc}),  decreases from $\delta t=534$ to 799 days, and increases again in the final epoch at $\delta t = 841$ days.

We note that our assumed SED shape is consistent with the X-ray non-detection reported in \citet{onori_case_2025}. If we extrapolate our first SED ($\delta t=197$ days, which is during the X-ray non-detection period) to the X-ray regime ($\sim1$ keV) with a power law $F\propto\nu^{-1}$, we expect a flux far lower ($2.5\times10^{-8}$ mJy) than the upper-limit of $3\times10^{-5}$ mJy, assuming a galactic hydrogen column density of $4.8\times10^{20} \rm cm^{-2}$ \citep{2016_HI4PI} and an X-ray photon index of $\Gamma_{\rm X}\sim2$.

\subsection{Cooling Break} \label{subsec:cooling}

Our fit to the final SED at $\delta t=841$--866 slightly over-predicts the highest frequency ($\gtrsim11$\,GHz) observations. Here, we consider whether this  discrepancy 
may be caused by the presence of a cooling break within the radio band. To assess this possibility, we fit the data with the SED model described in \citet{granot_shape_2002}:

\begin{dmath}
    F_{\nu_c} = F_\nu \left[1 + \left(\frac{\nu}{\nu_c}\right)^{s(\beta_2-\beta_3)}\right]^{-1/s} .
    \label{eq:eqc}
\end{dmath}

where $F_\nu$ is the model in Equation \ref{eq:eq1}, $\beta_3=-p/2$ is the slope above $\nu_c$, and $s=0.8-0.03p$ is the smoothing. Our best-fit model gives $\nu_c\approx7.4$ GHz, which is inconsistent with the calculated values given in Appendix \ref{app:selfcons}, where it is shown that $\nu_c$ is expected to rise through time, and is additionally outside of the data range for all possible values of $\epsilon_B$. Given this, the fact that the presumed break only appears in the final epoch, and the larger error bars on the final four data points in this epoch, we do not attribute the poor fit in the final epoch to the presence of a cooling break and proceed with a single break model.\footnote{While fitting for a cooling break does change the values of $\nu_p$ and $F_p$, we find that it does not change our overall results by a significant amount ($\sim 30\%$ for all equipartition quantities).}

\begin{figure*}[ht!]
\centering
\includegraphics[width=\linewidth]{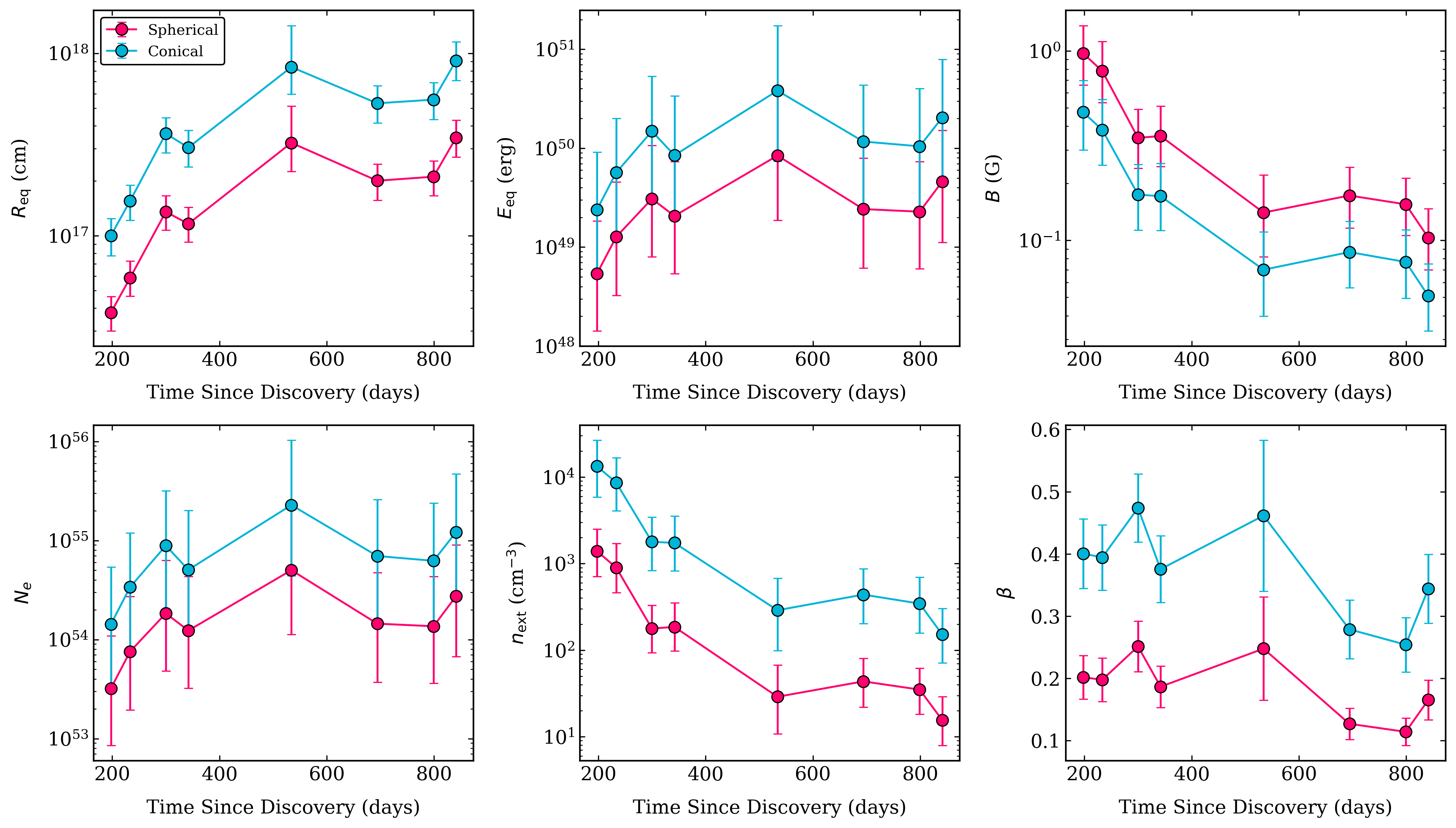}
\caption{
Evolution of the physical properties derived in an equipartition analysis of our SED best-fit parameters. Each subplot includes the results of the spherical geometry (pink) and conical geometry (blue). On the top row we show the emitting radius, outflow energy, and magnetic field strength as a function of time in the rest frame. The bottom row displays the number of radiating electrons, number density of electrons, and outflow velocity as a function of time.}
\label{fig:equipartition}
\end{figure*}

\section{Outflow Modeling} \label{sec:omodeling}

\subsection{Equipartition Analysis} \label{subsec:equipartition}
By estimating the peak frequency $\nu_p$, peak flux density $F_p$, and power-law index $p$, we can estimate the physical properties of the outflow. In this model, we assume the ambient electrons are accelerated into a power-law distribution of energies by the outflow (although, see the discussion of the impact of thermal electrons, or lack thereof, in Appendix \ref{app:thermal}), given by $N(\gamma) \propto \gamma^{-p}$, where $\gamma$ is the electron Lorentz factor, and $\gamma>\gamma_m$, where $\gamma_m$ is the minimum electron Lorentz factor. We consider $\nu_a \approx \nu_p$ and $\nu_m<\nu_a$. Following the procedures described in \citet{matsumoto_generalized_2023}, which builds upon the prescription of \citet{barniol_duran_radius_2013}, we assume that the electron and magnetic field energy densities are in equipartition to obtain a lower limit on the energy ($E_{eq}$) and a robust estimate of the radius of the emitting region ($R_{eq}$)\footnote{We note that this differs from other treatments, e.g. \cite{chevalier_synchrotron_1998}, also frequently used in the literature.}. We employ the equations derived in Rohde et al.\ (in prep.), a simultaneous application of additional $p$-dependent adjustments \citep{Shen_Zhang_2009} to the equipartition formalism presented in \citet{barniol_duran_radius_2013} and \citet{matsumoto_generalized_2023}, including hot proton corrections and electrons radiating at $\nu_m$.

We investigate two different geometries: a spherical and a conical emitting region. Both models are non-relativistic; the low luminosity of AT\,2022wtn ($\nu L_{\nu} \lesssim 10^{40}$ erg/s), combined with the lack of detected X-ray or $\gamma$-ray emission makes an on-axis relativistic jet unlikely. 

\begin{figure*}[ht!]
\centering
\includegraphics[width=\linewidth]{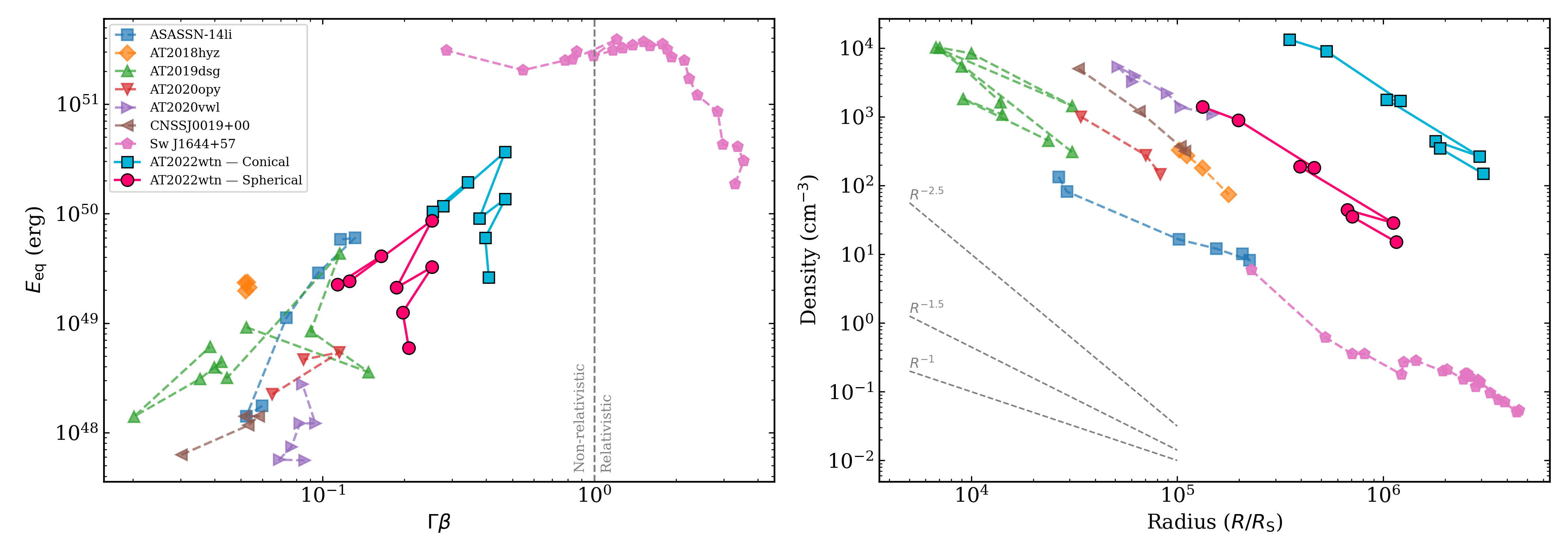}
\caption{
Left: The outflow kinetic energy and velocity of several previously studied TDEs; AT\,2022wtn is shown with circles and squares. Solutions for both the spherical and conical geometries are displayed in the same color scheme as Figure \ref{fig:equipartition}. Right: Circumnuclear density profile for the host galaxy of known TDEs, with the emitting radius being scaled by the SMBH Schwarzschild radius ($R_s=2GM/c^2$). AT\,2022wtn is displayed with the same visual scheme as in the left plot. We use a black hole mass of $M=10^6M_{\odot}$ for AT\,2022wtn, following \citet{onori_case_2025}. For AT\,2019dsg, when running our equipartition analysis we use a value of $\epsilon_{\rm B}=0.02$ following \citet{cendes_radio_2021}. The data and black hole masses are from ASASSN-14li \citep{alexander_discovery_2016}, AT\,2018hyz \citep{cendes_continued_2025}, AT\,2019dsg \citep{cendes_ubiquitous_2024}, AT\,2020opy \citep{goodwin_2020opy}, AT\,2020vwl \citep{goodwin_second_2025}, CNSSJ0019+00 \citep{anderson_caltechnrao_2020}, and Swift J1644+57 \citep{zauderer_birth_2011,eftekhari_radio_2018}.}
\label{fig:EvsBeta}
\end{figure*}

We characterize the geometry of the emitting region with standard area and volume filling factors  \citep{barniol_duran_radius_2013, matsumoto_generalized_2023}: $f_A=A/(\pi R^2/\Gamma^2)$ and $f_V=V/(\pi R^3/\Gamma^4)$, respectively. Here, $\Gamma$ is the Lorentz factor, which we set to $\Gamma=1$ for an assumed non-relativistic outflow. We assume that the emitting region is confined within a shell of radius 0.1$R_{eq}$, such that $f_A =1$ and $f_V=\frac{4}{3}(1-0.9^3) \approx 0.36$ in the spherical case \citep{alexander_discovery_2016,cendes_radio_2021}. For the conical geometry, we define $f_A =0.1$ to account for the effects of a mildly collimated outflow with an opening angle of $\sim20\degree$. Given this, we can derive the radius $R_{eq}$ and energy $E_{eq}$ (Rohde et al., in prep.)\footnote{These equations recover those provided in \citet{barniol_duran_radius_2013} when considering the Newtonian regime. They also recover the equations in \citet{matsumoto_generalized_2023} for a relativistic jet when $\xi=1$, $\epsilon=1$, and $p = 2$.}:

\begin{scriptsize}
\begin{equation}
\begin{split}
R_{eq} = \Bigg[
& \xi\,\epsilon\,\gamma_m^{2-p}
\frac{(p+1)\,\mathscr{C}^{p+5}\,c\,F_p^{p+6}\,
d_L^{2(p+6)}\,\eta^{\frac{5}{3}(p+5)}}%
{2^{p+2}\,11\sqrt{3}\,\pi^{p+7}\,m_e^{p+6}\,
\nu_p^{2p+13}\,(1+z)^{3p+19}} \\
& \times \frac{1}{f_A^{p+5}\,f_V}
\Bigg]^{\!\frac{1}{2p+13}}
\end{split}
\end{equation}

\begin{equation}
\begin{split}
E_{eq} = \Bigg[
& \xi^{11}\,\gamma_m^{11(2-p)}\,
\frac{17^{2p+13}\,\pi^{p+1}\,
\mathscr{C}^{3(p+1)}\,c^{4p+37}\,
m_e^{p+12}}
{2^{7p-4}\,(p+1)^{2(p+1)}\,3^5\,11^{11}\sqrt{3}} \\
& \times
\frac{F_p^{3p+14}\,d_L^{2(3p+14)}\,\eta^{5(p+1)}\,f_V^{2(p+1)}}
{q_e^{2(2p+13)}\,\nu_p^{2p+13}\,(1+z)^{5p+27}\,
f_A^{3(p+1)}} 
\Bigg]^{\!\frac{1}{2p+13}} \\
& \times
\left(
\frac{11}{17}\epsilon^{-\frac{2(p+1)}{2p+13}} +
\frac{2(p+1)}{17}\epsilon^{\frac{11}{2p+13}}
\right)
\end{split}
\end{equation}
\end{scriptsize}

\noindent where $\gamma_m = 2$ for non-relativistic sources (see also Appendix B), $d_L$ is the luminosity distance per $10^{28}$cm, and $z$ is the redshift of the system. $\mathscr{C}$ is a $p$-dependent correction factor that relates the synchrotron and Rayleigh-Jeans peak defined as \citep{Shen_Zhang_2009},
\begin{equation}
    \shenzhangfactor
    \label{eq:curlyc}
\end{equation}
where $G$ is the gamma function. $\xi$ accounts for hot proton corrections, and $\epsilon$ parameterizes the deviation from equipartition:

\begin{equation}
\xi \equiv 1+\epsilon_e^{-1}, \qquad
\epsilon \equiv \frac{11}{2(p + 1)} \frac{\epsilon_B}{\xi \epsilon_e}. \label{eq:epsB}
\end{equation}

To estimate the minimum energy of the system, we set $\epsilon =1$. We further assume that the fraction of energy carried by the electrons is $10\%$, $\epsilon_e = 0.1$ (for discussion on $\epsilon_B$, see Appendix B, and for a discussion of the impact of thermal electron emission, see Appendix \ref{app:thermal}). 

With $R_{eq}$ and $E_{eq}$, we can find the magnetic field strength $B$ and the number of electrons within the emitting region $N_e$ \citep{barniol_duran_radius_2013,matsumoto_generalized_2023}. Knowing $R_{eq}$ and $N_e$ allows us to calculate the number density of ambient electrons in the circumnuclear environment, $n_{\rm ext}=3N_e/f_\Omega\pi R_{eq}^3$\footnote{We don’t utilize a shock compression factor of 1/4 in our calulation of $n_{\rm ext}$ because the density of the CNM is calculated directly by dividing $N_e$ by the volume of the CNM swept-up by the shock; $f_\Omega$ is a geometric factor that correctly accounts for the total swept up volume \citep[see][]{matsumoto_generalized_2023}. Other works \citep[e.g.,][]{christy_peculiar_2024,cendes_radio_2021} estimate the number density of the electrons in the outflow first, which then requires an additional factor of 1/4 to reach the number density of electrons in the ambient medium.}. $f_\Omega$ is a solid-angle-filling factor that we set to 4 in the spherical case, and 0.1 in the conical case\footnote{$f_\Omega$ creates a factor of 40 difference in $n_{\rm ext}$ between the two models, which leads to a higher number density for the conical case. We note that this differs from previous analyses that consider both of these geometries (e.g., \citealp{alexander_discovery_2016,goodwin_at2019azh_2022}), where the conical density is lower, and $f_\Omega$ is not considered.}. The temporal evolution of the physical parameters for both geometries are shown in Figure \ref{fig:equipartition} and Table \ref{tab:equipartition}.

To validate this equipartition model we recompute the values derived in the literature for other non-relativistic TDEs: ASASSN-14li \citep{alexander_discovery_2016}, AT\,2018hyz \citep{cendes_mildly_2022}, AT\,2019dsg \citep{cendes_radio_2021}, AT\,2020opy \citep{goodwin_2020opy}, and CNSSJ0019+00 \citep{horesh_delayed_2021}. We use parameters following from the respective papers (e.g., $\epsilon_B = 0.1$ except for AT\,2019dsg which has $\epsilon_B = 0.02$). The results of this are shown in Figure \ref{fig:EvsBeta}. The values we find are broadly consistent with the previous equipartition results. Minor disagreements can be traced to the differences in assumptions made here as compared to other works (e.g., using a p-dependent $\mathscr{C}$ rather than setting $\mathscr{C} = 3$, as is done in \citealp{barniol_duran_radius_2013}).

\subsection{Launch Date} \label{subsec:launchdate}

We find the velocity of the outflow by rearranging equation 22 in \citet{barniol_duran_radius_2013} to solve for $\beta$:

\begin{dmath}
    \beta=\left[\frac{ct}{R_{eq}(1+z)}+1\right]^{-1},
\end{dmath}

where $t$ is the time since the outflow launch. As mentioned in Section \ref{subsec:lightcurve}, the Ku-band luminosity displays a genuine rise in flux between $\delta t = 167-233$ days, consistent with the launch of a new outflow. By performing a linear fit on $R_{eq}$, we can extrapolate back to $R_{eq}=0$ to infer a launch date; however, this requires us to assume that the outflow has remained in free expansion (i.e., a constant velocity) over the range of radio observations ($\delta t=197-866$d). Fitting a simple linear function to the first five epochs, where the radius is still evolving linearly, gives a launch date of $\delta t = 138\pm5$ days. A broken power-law (e.g., $R\propto t^1 \rightarrow R\propto t^{2/5}$ for a Sedov-Taylor-like expansion) also fits the data, but the lack of early-time radio coverage and the introduction of a new parameter ($t_{\rm break}$) do not allow for a robust constraint on a launch date due to degeneracy between parameters. 

If we instead fit a power law to the data, we find $R\propto t^{0.53}$, and a launch date of $\delta t\approx 179$ days, which is comparable to, but after the first radio detection at $\delta t = 167$ days. An $R\propto t^1$ fit to the first three data points yields an outflow launch date of $\delta t\approx156$ days; while technically permissible, we consider such a late launch date unlikely given that we measured the first radio detection just 11 days later. Given the uncertainty in potential launch dates, we assume that the outflow was in free-expansion for the first five epochs and was launched $\delta t \approx 138$ days after optical discovery, following the methods used in other TDE literature (e.g., \citealp{alexander_discovery_2016, christy_dichotomy_2025}).

\subsection{Earlier Outflow?} \label{subsec:otheroutflow}

In Section \ref{subsec:seds}, we excluded the data points corresponding to an excess transient component from our modeling of the first SED at $\delta t = 197$ days. If we instead take these points as comprising a distinct SED  from an entirely separate outflow launched at $\leq197$ days, we can use the lowest frequency detection as an upper limit on $\nu_p$ and a lower limit on $F_p$ of this component. We can then calculate all of the relevant equipartition quantities, obtaining a lower limit on the radius and energy of the outflow (which are time-independent in the Newtonian regime). These values are extremely high with respect to the values seen in the $\delta t=197$ days epoch, suggesting that if this is indeed a separate (perhaps prompt) outflow, it would have been moving extremely fast with a velocity of $v>0.3c$, even with a launch date of $\delta t=0$ days. Given the non-detection at $\delta t=97$ days, it is unlikely that an extremely fast and energetic outflow was launched before this time, implying an even higher velocity. While the implications of such an outflow are interesting, it is very difficult to draw any conclusions on it given the paucity of data, since this component disappears almost completely in the second SED at 233 days. We therefore opt to exclude it from the modeling, instead focusing on the post-138 days outflow in our analysis.

\section{Discussion} \label{sec:discussion}

Using the results obtained in our equipartition analysis, we will now discuss the possible outflow mechanisms behind AT\,2022wtn, and compare its physical properties to that of other TDEs. 

\subsection{Outflow Velocity and Kinetic Energy} \label{subsec:energyvelocity}

We plot the equipartition energy as a function of outflow velocity for AT\,2022wtn and several other TDEs in Figure \ref{fig:EvsBeta}. To standardize our comparison, we recompute the equipartition values for these other TDEs following the method described in Section \ref{subsec:equipartition}. Our non-relativistic spherical model predicts a roughly constant velocity for the first five epochs ($\delta t=197-534$d), before decelerating in the final three ($\delta t\gtrsim534$d; see Figure \ref{fig:equipartition}). This corresponds to $\beta\approx0.21\pm0.03$ for a spherical geometry, and $\beta\approx0.41\pm0.04$ for a conical geometry. These values are quite high in the context of other thermal TDEs, which usually have velocities of $\beta\approx0.05-0.1$ \citep{alexander_radio_2020, alexander_discovery_2016, christy_dichotomy_2025, goodwin_at2019azh_2022}.  When modeled as a spherical outflow, AT\,2018hyz displayed velocities of $\beta\approx0.33$ and an energy comparable to that of AT\,2022wtn \citep{cendes_continued_2025}; however, there is strong evidence that AT\,2018hyz possesses an off-axis relativistic jet, which is not the case for AT\,2022wtn (see Section \ref{subsubsec:jet}).

AT\,2022wtn exhibits a higher kinetic energy than any other thermal (i.e., non-relativistic) TDE (see Figure \ref{fig:EvsBeta}; \citealp{Mockler_2021}), but does not reach the energies seen in the relativistic on-axis TDEs Swift J1644+57 and AT\,2022cmc \citep{eftekhari_radio_2018,2023_2022cmc}. Furthermore, our inferred equipartition energy increases by roughly $\sim5\times$ between $\delta t = 197 - 300$ days, driven by the increase in peak flux (see Figure \ref{fig:peakvalues}), at which point it plateaus at $\sim3.7\times10^{49}$ erg. The conical model sees a similar evolution, but with a higher energy at all times, plateauing at $\sim1.8\times10^{50}$ erg. Such an increase in energy has previously been observed in TDEs \citep[e.g.,][]{goodwin_at2019azh_2022,cendes_radio_2021,berger_radio_2012,christy_peculiar_2024}, and could be caused by energy injection into the outflow due to accretion onto the SMBH \citep{cendes_radio_2021}. An energy increase could also be due to a range of ejecta velocities present in the outflow \citep{berger_radio_2012}, but this is unlikely in our case since the expected contemporaneous decline in velocity is not seen  (see Figure \ref{fig:equipartition}). 

\subsection{Circumnuclear Density Profile} \label{subsec:densityprofile}

Figure \ref{fig:EvsBeta} shows the circumnuclear density $n_{ext}$ as a function of the radius of the emitting region for AT\,2022wtn compared to other TDEs in the literature. The density profile for the spherical case follows a slope of $n_{ext}\propto R^{-2.08}$ throughout the observation period. This falls within the range of other non-relativistic TDEs, some of which display a profile of $n_{ext}\propto R^{-1.5}$ consistent with spherical Bondi accretion \citep{1952MNRAS.112..195B}, and others displaying a steeper $n_{ext}\propto R^{-2.5}$ profile \citep{alexander_discovery_2016, anderson_caltechnrao_2020}. 

\subsection{Outflow Mechanism} \label{subsec:emission}

Armed with the extensive radio analysis presented thus far, we can now discuss implications for the outflow mechanism behind AT\,2022wtn. We first consider a relativistic jet scenario before looking at several non-relativistic outflow models, including unbound debris, collisionally-induced outflows, accretion-driven wind, and an outflow driven by a state transition in the accretion disk. We analyze these scenarios in the context of the multiwavelength data presented in \citet{onori_case_2025}, including their discovery of a prompt accretion disk formation and presence of outflows at early times, and their calculation of the Eddington ratio. 

\subsubsection{Relativistic Jet} \label{subsubsec:jet}

So far, we have assumed that the outflow from AT\,2022wtn is non-relativistic. Instead, if we assume a collimated relativistic outflow ($\Gamma\approx2$; \citealp{alexander_discovery_2016,goodwin_at2019azh_2022}), we obtain $f_A \lesssim10^{-5}$ and an opening angle of $\theta_j\lesssim0.01\degree$ by solving for $f_A$ with equation 24 in \citet{barniol_duran_radius_2013}. This is far narrower than typical jets, even in GRBs \citep{Frail_2001}. In addition, if the radio emission from AT\,2022wtn were caused by a relativistic jet, we would expect to see energies and radio luminosities  similar to that of Swift J1644+57 ($\sim10^{52}$ erg and $\nu L_{\nu} \gtrsim 10^{40}$ erg/s; \citealp{eftekhari_radio_2018}), which are not observed. 

An on-axis relativistic jet is further ruled out by considering the multiwavelength properties of AT\,2022wtn, as outlined in \citet{onori_case_2025}. Due to the strong Bowen lines present in the optical and UV spectra, and the lack of X-ray emission, it is likely that we are seeing the accretion disk from an off-axis viewing angle ($\theta>0\degree$). However, for the high energies ($\sim10^{50}$ erg) seen in AT\,2022wtn, we would expect a more extended light curve (i.e., a slower evolution) than we observe (see right panel in Figure \ref{fig:fig1and2}) if the radio emission instead arose from a promptly-launched off-axis jet \citep{granot_off-axis_2018}.


In summary, the evolution of the light curve, the low luminosity and energy, and the unphysical opening angle make the possibility of a relativistic jet being the source of the observed radio emission in AT\,2022wtn unlikely. 

\subsubsection{Unbound Debris} \label{subsubsec:unbound}

During a TDE, roughly half of the debris from the disrupted star will be gravitationally unbound from the SMBH, escaping at a speed dependent on the ratio of the SMBH's tidal radius and the pericenter distance \citep{yalinewich_radio_2019}. This debris can form a bow shock as it interacts with the CNM, producing synchrotron emission. The unbound material is expected to possess energies of $\sim 10^{47}-10^{48}$ erg \citep{krolik_asassn-14li_2016} and travel at speeds of $\lesssim0.06c$ \citep{yalinewich_radio_2019}. 

The energy found in our spherical analysis of $\sim3.7\times10^{49}$ erg exceeds the allowed range expected for an unbound debris stream. Additionally, our inferred velocity ($\beta\approx0.21$) is far higher than expected in an unbound debris model. This model also requires a collimation ($f_A\sim0.2$) higher than in the spherical case; increasing the collimation would only further increase the energy and velocity seen in our analysis. Therefore, an unbound debris stream causing the radio emission is unlikely. 

\subsubsection{Collisionally-induced Outflow} \label{subsubsec:CIO}

\citet{lu_self-intersection_2020} demonstrated that stream-stream collisions in the bound stellar debris can cause a spherical outflow of gas, referred to as a collisionally-induced outflow (CIO). The expected energy in a CIO model is $\sim10^{50}$ erg, consistent with the values obtained in our equipartition analysis. 

However, the velocity we derive for a spherical geometry ($\beta\approx 0.21$) exceeds the expected values for this model ($\beta\approx0.01-0.1$). In addition, the CIO scenario predicts that a radio-emitting outflow would be launched around the time of optical discovery, as the optical emission in this model is primarily driven by stream-stream interactions \citep{lu_self-intersection_2020}. This is disfavored by our inferred launch date of $\delta t =138$ days, and the radio upper limit at $\delta t=97$ days.

\subsubsection{Accretion-Driven Wind} \label{subsubsec:accwind}

The radiation pressure generated by SMBH accretion can produce winds traveling at velocities of $\beta\approx0.01-0.1$; this would generate radio emission as the winds shock the surrounding CNM \citep{2009MNRAS.400.2070S}. TDE outflows have been explained with this mechanism before \citep{alexander_discovery_2016,cendes_radio_2021}, but the outflow velocity of AT\,2022wtn ($\beta\approx0.21$) is inconsistent with the expected values ($\beta\approx0.01-0.1$) in an accretion wind scenario \citep{2009MNRAS.400.2070S}. 

There is evidence for a spherical outflow found in the multiwavelength data presented by \citet{onori_case_2025}; however, this outflow would be launched near the time of optical discovery, far earlier than our proposed launch date of $\delta t=138$ days; it would also be much slower than our expected velocity, only reaching $v\approx0.007c$. Further evidence for an outflow is found at $\delta t \approx 80$ days, where broad optical spectral lines are detected. However, the emission lines seen in this outflow are still much narrower than would be expected for an outflow with $v\approx0.21c$ \citep{onori_case_2025}, indicating that this is a separate component not related to the radio emission.

\subsubsection{Outflow Driven by an Accretion State Transition} \label{subsubsec:subjet}

We finally consider the possibility that the radio emission is caused by a delayed, sub-relativistic outflow resulting from a state change in the accretion disk, first proposed in \citet{De_Colle_2012}. The outflow could be approximated with either of our equipartition models; a conical geometry would display a higher energy, velocity, and radius than in the spherical case. Such a scenario is consistent with what is proposed in \citet{wu2025delayedradioemissiontidal}, where instabilities or state transitions within the accretion disk lead to the launch of an outflow with velocities of $0.05c-0.3c$, in line with our inferred value of $\beta\approx0.21$ for the spherical case. The predicted energy of $\sim 10^{50}$ in this model is consistent with our values ($\sim3.7\times10^{49}$, spherical; $\sim1.8\times10^{50}$, conical) as well. 

This model is attractive because of its ability to explain the delayed radio emission seen in AT\,2022wtn. \citet{onori_case_2025} find evidence for a prompt circularization process, far earlier than the first radio detection at $\delta t = 167$ days; if the accretion flow were to undergo a state transition (e.g., into a delayed phase of super-Eddington accretion) around our inferred launch date, we would expect an outflow to be launched at roughly the same time \citep{2025_alexander_review}. This state transition could be caused by the promptly-circularized debris forming a hot pressure-supported envelope, which would delay rapid accretion until the envelope cools \citep{Metzger_2022}, as suggested in \citet{2025_alexander_review}. Another possible explanation for this delayed phase of accretion is a larger debris self-intersection radius; in this case, the debris would take longer to reach the event horizon, delaying peak accretion onto the SMBH \citep{nicholl_systematic_2022}.

Such a transition into super-Eddington accretion is consistent with the findings in \citet{onori_case_2025}, who report an Eddington ratio of $0.06 - 0.09$ (i.e. sub-Eddington) for $\delta t \lesssim 120$ days. Additionally, if the radio emission were produced by an outflow launched due to a new phase of super-Eddington accretion, the rapid rise of $\lesssim100$ days in the light curve of AT\,2022wtn (see Figure \ref{fig:fig1and2}) would be expected, as the shock collides with the surrounding CNM \citep{wu2025delayedradioemissiontidal}. We therefore conclude that a transition into a phase of delayed super-Eddington accretion can explain the outflow in AT\,2022wtn. 

\section{Conclusions} \label{sec:conclusion}

We present the results of our detailed study of the TDE AT\,2022wtn, observed in the radio from $\delta t =97-866$ days after its optical discovery. Transient radio emission was first observed several months after the peak of the optical flare, rising to a peak flux density at $\delta t = 300$ days, before beginning to decline. The radio luminosity of AT\,2022wtn is comparable to that of other thermal TDEs (see Figure \ref{fig:fig1and2}).

In our analysis, we modeled the radio emission with two distinct non-relativistic geometries: a spherical and conical outflow launched at $\delta t =138$ days after optical discovery. We find that the energy displays a $\sim5\times$ increase between $\delta t = 197-300$ days, possibly indicative of energy injection from accretion onto the central SMBH. The energy then plateaus at $\sim3.7\times10^{49}$ erg, a value higher than most other thermal TDEs, but still far lower than the energy expected in a relativistic jet (e.g., Swift 1644+57; \citealp{eftekhari_radio_2018}). We rule out the possibility of the radio emission originating in a relativistic jet, as the opening angle of the jet would have to be extremely small ($\lesssim0.1\degree$); additionally, AT\,2022wtn displays a lower luminosity than jetted-TDE candidates such as Swift J1644+57 and AT\,2018hyz \citep{eftekhari_radio_2018,cendes_continued_2025}.

We find that the outflow propagates at $v\approx0.21c$ in the spherical case, and $v\approx0.41c$ in the conical case, 
which is far too fast to be explained by several of the current non-relativistic TDE models; unbound debris streams, collisionally-induced outflows, and accretion-driven winds all require a velocity of $v\lesssim 0.1c$. A delayed outflow caused by an accretion state transition, which would have similar dynamical properties to either of our equipartition model geometries, offers a viable explanation for the observed radio emission. This model additionally explains the delayed launch date of $\delta t \approx138$ days, and the fast rise in the Ku-band light curve (see Figure \ref{fig:fig1and2}). In summary, we favor an outflow caused by an accretion state transition as the origin of the radio emission. 

AT\,2022wtn possesses a uniquely fast and energetic outflow, more powerful than that of other non-relativistic TDEs. Our analysis provides insight into the outflow properties of this event, and also emphasizes the diversity of the non-relativistic radio-emitting TDE population. We encourage continued future modeling that considers powerful and evolving outflows, and we also highlight the need for extended radio and multiwavelength monitoring campaigns to constrain the mechanisms behind TDE outflows.

\section*{Acknowledgments} \label{sec:acknowledgments}

The National Radio Astronomy Observatory and Green Bank Observatory are facilities of the U.S. National Science Foundation operated under cooperative agreement by Associated Universities, Inc. We thank the staff of the GMRT that made these observations possible. GMRT is run by the National Centre for Radio Astrophysics of the Tata Institute of Fundamental Research.

GF acknowledges support provided by the National Science Foundation Research Experience for Undergraduates award No.~2349237. AJG is grateful for support from the Forrest Research Foundation. NF acknowledges support from the National Science Foundation Graduate Research Fellowship Program under Grant No.~DGE-2137419. The authors thank E. Ramirez-Ruiz for helpful comments and suggestions. 

This work made use of the following software packages: \texttt{astropy} \citep{astropy:2013,astropy:2018,astropy:2022}, \texttt{Jupyter} \citep{2007CSE.....9c..21P,kluyver2016jupyter}, \texttt{matplotlib} \citep{Hunter:2007}, \texttt{numpy} \citep{numpy}, \texttt{pandas} \citep{mckinney-proc-scipy-2010,pandas_17806077}, \texttt{python} \citep{python}, \texttt{scipy} \citep{2020SciPy-NMeth,scipy_17873309}, and \texttt{OTTER} \citep{franz_python_2026, franz_open_2026}.

Software citation information aggregated using \texttt{\href{https://www.tomwagg.com/software-citation-station/}{The Software Citation Station}} \citep{software-citation-station-paper,software-citation-station-zenodo}.

\section*{Modeling Code} \label{sec:modelingcode}

The equipartition modeling software used in this work is publicly available on GitHub here: \url{https://github.com/rohdog2003/equipartition} \citep{rohde_equipartition_2026}. 

\appendix
\section{Flux density measurements} \label{app:fluxdensity}

Table \ref{tab:obsA} displays the flux density measurements for each observation epoch of AT\,2022wtn at each observing frequency. We report uncertainties as a $1\sigma$ statistical error alongside a 5\% systematic error, as discussed in Section \ref{sec:observations}. Non-detections are reported as $3\sigma$ upper limits.

\renewcommand{\thetable}{A1}
\setlength{\tabcolsep}{20pt} 
\begin{longtable}{lccccc}
\caption{Radio observations of AT\,2022wtn.} \label{tab:obsA} \\
\hline \hline
Date (UTC) & $\delta t$ (days) & Telescope & Configuration & $\nu$ (GHz) & $F_\nu$ ($\mu$Jy) \\
 &  &  &  &  & $\pm$ Statistical Error \\
  &  &  &  &  & $\pm$ Systematic Error \\
\hline
\endfirsthead
\hline
Date (UTC) & $\delta t$ (days) & Telescope & Configuration & $\nu$ (GHz) & $F_\nu$ ($\mu$Jy) \\
 &  &  &  &  & $\pm$ Statistical Error \\
  &  &  &  &  & $\pm$ Systematic Error \\
\hline
\endhead
\endfoot
\endlastfoot

2023-01-07 & 97 & VLA & C$\rightarrow$B & 15 & $<54$ \\
\hline
2023-03-20 & 169 & VLA & B & 15 & $202 \pm 9 \pm 10$ \\
\hline
2023-04-17 & 197 & VLA & B & 1.3 & $497 \pm 70 \pm 25$ \\
            &     &     &   & 1.8 & $291 \pm 32 \pm 15$ \\
            &     &     &   & 2.5 & $328 \pm 34 \pm 16$ \\
            &     &     &   & 3.5 & $437 \pm 22 \pm 22$ \\
            &     &     &   & 5 & $642 \pm 24 \pm 32$ \\
            &     &     &   & 7 & $816 \pm 23 \pm 41$ \\
            &     &     &   & 9 & $878 \pm 22 \pm 44$ \\
            &     &     &   & 11 & $907 \pm 28 \pm 45$ \\
            &     &     &   & 13.5 & $861 \pm 18 \pm 43$ \\
            &     &     &   & 16.6 & $769 \pm 20 \pm 38$ \\
            &     &     &   & 24 & $630 \pm 22 \pm 32$ \\
            &     &     &   & 20 & $728 \pm 21 \pm 36$ \\
\hline
2023-05-23 & 233 & VLA & B & 1.3 & $227 \pm 72 \pm 11$ \\
            &     &     &   & 1.8 & $237 \pm 32 \pm 12$ \\
            &     &     &   & 2.5 & $446 \pm 36 \pm 22$ \\
            &     &     &   & 3.5 & $751 \pm 21 \pm 38$ \\
            &     &     &   & 5 & $1269 \pm 24 \pm 63$ \\
            &     &     &   & 7 & $1671 \pm 23 \pm 84$ \\
            &     &     &   & 9 & $1669 \pm 21 \pm 83$ \\
            &     &     &   & 11 & $1565 \pm 25 \pm 78$ \\
            &     &     &   & 13.5 & $1326 \pm 21 \pm 66$ \\
            &     &     &   & 16.6 & $1167 \pm 21 \pm 58$ \\
            &     &     &   & 24 & $786 \pm 27 \pm 39$ \\
            &     &     &   & 20 & $995 \pm 22 \pm 50$ \\
\hline
2023-07-29 & 300 & VLA & A & 1.3 & $552 \pm 36 \pm 28$ \\
            &     &     &   & 1.8 & $1049 \pm 33 \pm 52$ \\
            &     &     &   & 2.5 & $1579 \pm 26 \pm 79$ \\
            &     &     &   & 3.5 & $1872 \pm 21 \pm 94$ \\
            &     &     &   & 5 & $1794 \pm 22 \pm 90$ \\
            &     &     &   & 7 & $1433 \pm 20 \pm 72$ \\
            &     &     &   & 9 & $1117 \pm 20 \pm 56$ \\
            &     &     &   & 11 & $855 \pm 25 \pm 43$ \\
            &     &     &   & 13.5 & $562 \pm 19 \pm 28$ \\
            &     &     &   & 16.6 & $720 \pm 16 \pm 36$ \\
            &     &     &   & 24 & $315 \pm 37 \pm 16$ \\
            &     &     &   & 20 & $413 \pm 28 \pm 21$ \\
\hline
2023-09-09 & 342 & VLA & A & 1.3 & $315 \pm 30 \pm 16$ \\
            &     &     &   & 1.8 & $634 \pm 31 \pm 32$ \\
            &     &     &   & 2.5 & $1186 \pm 26 \pm 59$ \\
            &     &     &   & 3.5 & $1331 \pm 18 \pm 67$ \\
            &     &     &   & 5 & $1225 \pm 21 \pm 61$ \\
            &     &     &   & 7 & $1048 \pm 28 \pm 52$ \\
            &     &     &   & 9 & $753 \pm 19 \pm 38$ \\
            &     &     &   & 11 & $550 \pm 23 \pm 28$ \\
            &     &     &   & 13.5 & $369 \pm 16 \pm 18$ \\
            &     &     &   & 16.6 & $466 \pm 15 \pm 23$ \\
            &     &     &   & 24 & $217 \pm 24 \pm 11$ \\
            &     &     &   & 20 & $301 \pm 21 \pm 15$ \\
\hline
2024-03-19 & 534 & VLA & C & 2.5 & $1463 \pm 128 \pm 73$ \\
            &     &     &   & 3.5 & $1112 \pm 74 \pm 56$ \\
            &     &     &   & 5 & $787 \pm 33 \pm 39$ \\
            &     &     &   & 7 & $564 \pm 23 \pm 28$ \\
            &     &     &   & 9 & $429 \pm 18 \pm 21$ \\
            &     &     &   & 11 & $339 \pm 19 \pm 17$ \\
            &     &     &   & 13.5 & $264 \pm 16 \pm 13$ \\
            &     &     &   & 16.6 & $224 \pm 15 \pm 11$ \\
\hline
2024-08-26 & 694 & VLA & B & 1.52 & $806 \pm 44 \pm 40$ \\
            &     &     &   & 3 & $601 \pm 25 \pm 30$ \\
            &     &     &   & 5 & $413 \pm 25 \pm 21$ \\
            &     &     &   & 7 & $274 \pm 22 \pm 14$ \\
            &     &     &   & 9 & $227 \pm 21 \pm 11$ \\
            &     &     &   & 11 & $195 \pm 29 \pm 10$ \\
            &     &     &   & 13 & $162 \pm 30 \pm 8$ \\
            &     &     &   & 15 & $110 \pm 31 \pm 6$ \\
            &     &     &   & 17 & $85 \pm 38 \pm 4$ \\
\hline
2024-12-09 & 799 & VLA & A & 1.5 & $635 \pm 25 \pm 32$ \\
            &     &     &   & 3 & $423 \pm 12 \pm 21$ \\
            &     &     &   & 5 & $308 \pm 16 \pm 15$ \\
            &     &     &   & 7 & $194 \pm 13 \pm 10$ \\
            &     &     &   & 9 & $135 \pm 13 \pm 7$ \\
            &     &     &   & 11 & $103 \pm 15 \pm 5$ \\
            &     &     &   & 13 & $77 \pm 11 \pm 4$ \\
            &     &     &   & 15 & $80 \pm 11 \pm 4$ \\
            &     &     &   & 17 & $65 \pm 13 \pm 3$ \\
\hline
2025-01-20 & 841 & VLA & A & 1.5 & $757 \pm 26 \pm 38$ \\
            &     &     &   & 3 & $408 \pm 14 \pm 20$ \\
            &     &     &   & 5 & $223 \pm 15 \pm 11$ \\
            &     &     &   & 7 & $154 \pm 11 \pm 8$ \\
            &     &     &   & 9 & $106 \pm 12 \pm 5$ \\
            &     &     &   & 11 & $70 \pm 16 \pm 4$ \\
            &     &     &   & 13 & $53 \pm 10 \pm 3$ \\
            &     &     &   & 15 & $49 \pm 10 \pm 2$ \\
            &     &     &   & 17 & $37 \pm 13 \pm 2$ \\
2025-02-14 & 866 & GMRT & -- & 0.7 & $407 \pm 112 \pm 20$ \\
\hline
\end{longtable}

\onecolumngrid
\section{Constraints on Free-Free Absorption in The Early Radio Light Curve} \label{app:freefree}
The VLA Ku-band ($\approx15$\,GHz) observations at 97 days (upper limit) and 169 days and 197 days (detections) reveal a very rapid rise in the radio light curve of at least $\sim t^{2.4}$ (from 97--169 days) and $\sim t^9$ (from 169--197 days). As discussed in Section~\ref{subsec:launchdate}, one possibility is that this rise is associated with the delayed launch of an outflow. Here, we consider the alternate scenario where the outflow is launched earlier but the radio emission is initially suppressed by free-free absorption (FFA). The free-free optical depth at frequency $\nu$ through a path-length $L$ for a fully ionized medium is given by \citep{rl1986}
\begin{equation*}
    \tau_{\rm ff} = 1.8\times10^{-2} g_{\rm ff} (T/K)^{-1.5} Z^2 (n/{\rm cm}^{-3}) (L/{\rm cm)} (\nu/{\rm Hz})^{-2},
\end{equation*}
where we set the Gaunt factor, $g_{\rm ff}\sim5$ and ion charge, $Z=1$ (assuming pure hydrogen composition) following  \cite{sfaradi_first_2025} and \cite{nayana_at2024wpp}. Setting the radius to the equipartition value inferred from the first full SED at 197 days yields  
\begin{equation*}
    \tau_{\rm ff} = 0.5 (T/10^5 K)^{-1.5} (n/10^6{\rm cm}^{-3}) (L/{4\times10^{16}\rm cm)} (\nu/{15\rm GHz})^{-2}.
\end{equation*}
A suppression corresponding to the observed flux ratio of $e^{\tau_{\rm ff}}\approx4$ between subsequent pairs of the first three epochs corresponds to $\tau_{\rm ff}\approx1.4$, requiring an average density of $\approx2\times10^6{\rm cm}^{-3}$ within the absorbing region and an enclosed mass of $\approx0.4\,M_\odot$ within a spherical volume of radius $L$. This large enclosed mass hints that the suppression is unlikely to arise from FFA. Although this calculation remains subject to the unknown geometry of the system and the properties (e.g., temperature, size, and density structure) of the absorbing medium, yet more evidence against FFA suppression is afforded by the SEDs. If the Ku-band flux were subject to strong FFA, this suppression would increase rapidly at lower frequencies (as $e^{-1/\nu^2}$); however, the SED \textit{rises} below Ku-band at 197 days and there is no evidence for a steep (free-free absorbed) spectrum below the peak. The FFA model can be rescued if the FFA is assumed to entirely disappear by 197 days; however, the flux on the optically thin segment of the SED continues to rise from 197 days to 300 days while the spectrum remains peaked at lower frequencies (Fig.~\ref{fig:mcmc}). This is very difficult to explain as a further reduction in $\tau_{\rm ff}$, since the expected change in the spectral shape this should cause is not seen. Whereas we cannot rule out a combination of rapidly dropping $\tau_{\rm ff}$ and a rising flux due to, e.g., energy injection into the outflow, such scenarios appear increasingly contrived and a delayed outflow launch remains the simpler, favored interpretation. 

\section{Self-consistency checks for synchrotron Break Frequencies} \label{app:selfcons}

\begin{figure*}[ht!]
\centering
\includegraphics[width=\linewidth]{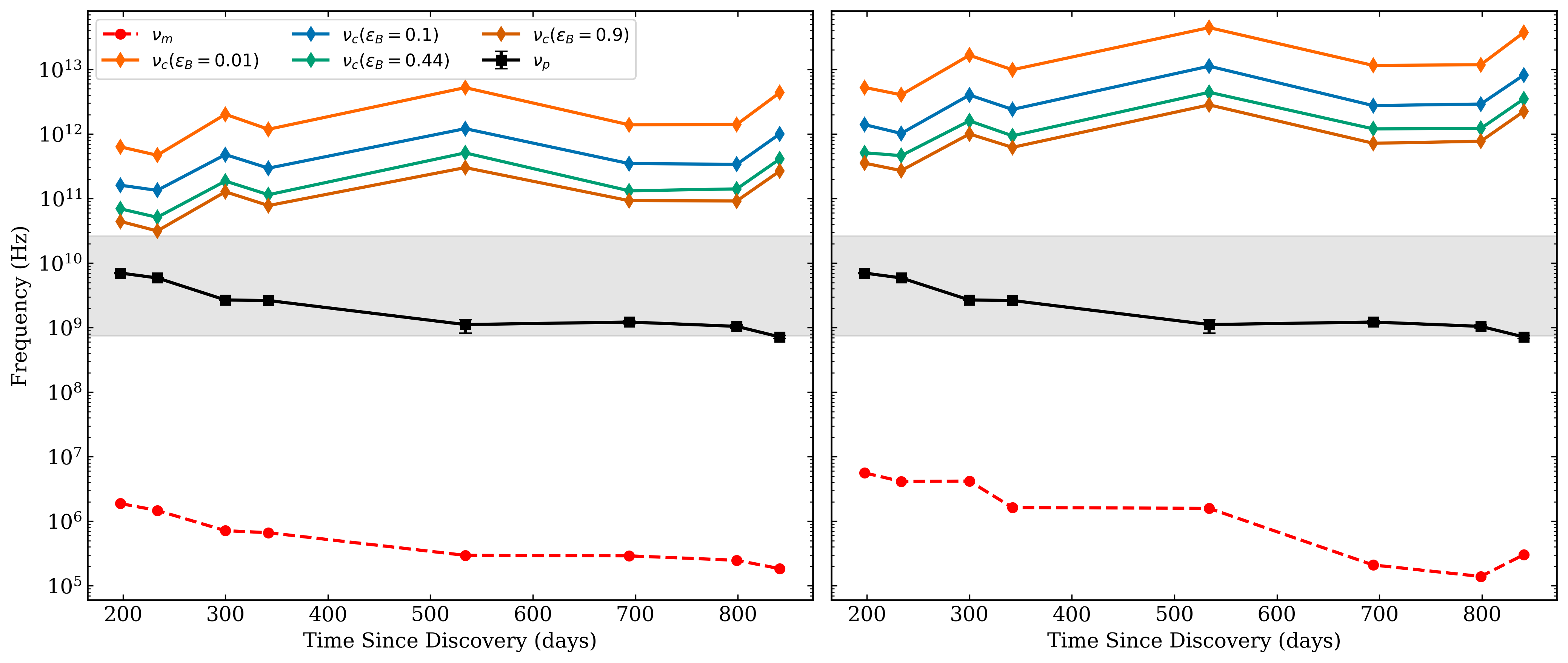}
\caption{
Calculated values for $\nu_m$ and $\nu_c$ (for several different values of $\epsilon_B$) plotted alongside the peak frequency $\nu_p$. The left plot shows results for the spherical geometry, and the right plot shows results for the conical geometry. The shaded region represents the range of frequency that our data covers.}
\label{fig:nucnum}
\end{figure*}

To assess the self-consistency of our choice of model in Section \ref{sec:emodeling} ($\nu_m < \nu_a < \nu_c$), we calculate the location of the break frequencies $\nu_m$ and $\nu_c$ with equation 6 in \citet{cendes_continued_2025} and equation 7 in \citet{cendes_mildly_2022}, respectively: 

\begin{equation}
\nu_m \approx 4.2 \times 10^{5}\ \text{Hz}
\left[
\frac{\nu_{p,10}^{3/4}}{\eta^{5/12}}
\left( \frac{t}{100\,\text{d}} \right)^{-1/4}
\right]
\frac{f_A^{1/4}}{f_V^{1/4}} \gamma_m^{2} \,,
\end{equation}

\begin{equation}
\nu_c \approx 2.25 \times 10^{14} B^{-3} t_d^{-2}\ \text{Hz}.
\end{equation}

where $\nu_{\rm p,10}=\nu_{\rm p}/10$ GHz, $\eta=\text{max}[1,\nu_{\rm m}/\nu_{\rm a}]$, $f_A$ and $f_V$ are the area and volume filling factors, and $\gamma_m=\max[\frac{9}{32} \mu (\frac{m_p}{m_e})\epsilon_e(\frac{p-2}{p-1})(\frac{v_{sh}}{c})^2,2]$, where $v_{sh}$ is the velocity of the shock, and we set $\mu=1$, assuming a pure hydrogen composition. B is the magnetic field derived from our equipartition analysis, and $t_d$ is the time since the outflow launched, in days. Our calculated values for both outflow geometries are shown in Figure \ref{fig:nucnum}. 

For both the spherical and conical case, it can be seen that the ordering follows $\nu_m < \nu_a < \nu_c$ if we take $\nu_p\approx\nu_a$. It is notable that $\nu_c>\nu_a$ for all times and values of $\epsilon_B$; this excludes the possibility of the electrons being in the fast cooling regime, where the spectral indices follow $\frac{5}{2} \rightarrow -\frac{p}{2}$. In both models, neither $\nu_c$ or $\nu_m$ pass within the data frequency range; this is consistent with how we modeled the SEDs in Section \ref{subsec:seds} as a single broken power law, with a break at $\nu_a$. 

\begin{figure}[ht!]
\centering
\includegraphics[width=0.5\linewidth]{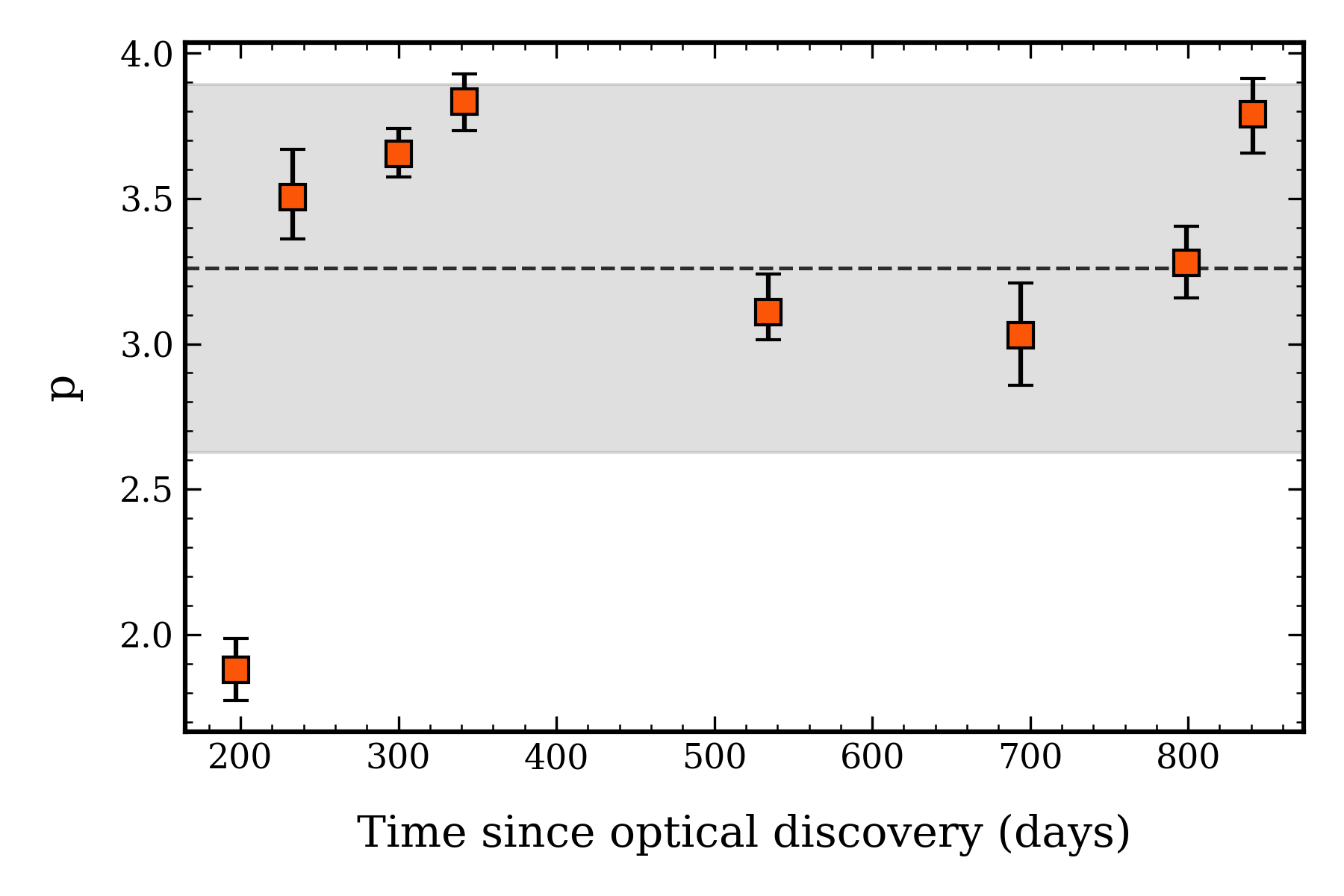}
\caption{
Evolution of the synchrotron energy index $p$ through time, when left as a free parameter in our MCMC modeling. The dashed line represents the average value and the shaded region is the standard deviation (across all epochs) of the per-epoch medians.}
\label{fig:pvalues}
\end{figure}

We also note that while the expression for $\gamma_m$ used here is consistent with the assumed $\gamma_m=2$ in Section \ref{sec:omodeling} for the spherical model (i.e. the calculated $\gamma_m=2$ for all epochs), for the conical case we infer a mean value of $\gamma_m\sim4$. This results in only a $\lesssim 5\%$ reduction in the $R_{\rm eq}$ but a $\approx 40\%$ reduction in $E_{\rm eq}$ relative to assuming $\gamma_m=2$. In the main text, we assume equipartition, and $\gamma_m=2$, primarily to compare AT\,2022wtn to other TDEs that have had the same assumptions made (e.g., \citealp{goodwin_at2019azh_2022,christy_dichotomy_2025}). This assumption is valid in the spherical case, but an accurate comparison for the conical case, using the calculated value of $\gamma_m$, would require us to remodel every other TDE within this framework (i.e., $\gamma_m\neq 2$) and is beyond the scope of this paper. In summary, we find that the calculated values of the break frequencies $\nu_m$ and $\nu_c$ are consistent with our assumptions in Section \ref{subsec:seds}, and that our calculated values of $\gamma_m$ are consistent with the assumptions made in Section \ref{sec:omodeling} for the spherical case, but some deviations are expected from $\gamma_m > 2$ in the conical case. 

\section{Evolving electron energy index} \label{app:evolvingp}

In Figure~\ref{fig:pvalues}, we present the results of fitting the SEDs leaving $p$ as a free parameter. We find a large change in $p$ from 197 days to 233 days, with modest variations in $p$ thereafter. These variations in $p$ would significantly impact the equipartition quantities, in particular, implying a very large energy increase by a factor of $\approx100$ between $\delta t=197$--233 days. While energy injection has been proposed for TDEs before (see Section~\ref{subsec:energyvelocity}), this extremely rapid energy increase appears implausible. Fitting all SEDs with a fixed optically thin spectral slope still yields good fits, suggesting that this variation is not strongly supported by the data. Finally, we find that this variation is substantially reduced on fixing the smoothing parameter to $s=1$ (which otherwise depends on $p$ in the \citealt{granot_shape_2002} framework) and is therefore likely not a real effect. To avoid imprinting this unphysical variation on the equipartition quantities, we undertake our analysis with $p$ fixed to the mean value of $3.26$. 

\defcitealias{margalit_peak_2024}{MQ24}
\newcommand{\mq}{\citetalias{margalit_peak_2024}}

\section{Constraints on emission from Thermal Electrons}\label{app:thermal}
For completeness, we explore the impact of thermal electrons on our SED modeling in \autoref{sec:omodeling} using the semi-analytic model of \citet[][hereafter \mq]{margalit_peak_2024}. This model accounts for synchrotron emission and absorption by thermal and non-thermal electrons for both relativistic and non-relativistic outflows. The \mq\ model has $\sim7$ primary free parameters: $p$, the synchrotron power law index; $\beta\Gamma$, the shock 4-velocity; $n$, the number density of the ambient medium; $\epsilon_T$, $\epsilon_e$, and $\epsilon_B$, the parameterization of the energy stored in the thermal electrons, relativistic electrons, and magnetic field respectively; and $\delta t_{\rm outflow}$, the time since outflow launch. With the exception of $\epsilon_T$, these parameters have similar meanings as described in \autoref{sec:omodeling}.

\begin{figure}
    \centering
    \includegraphics[width=0.95\linewidth]{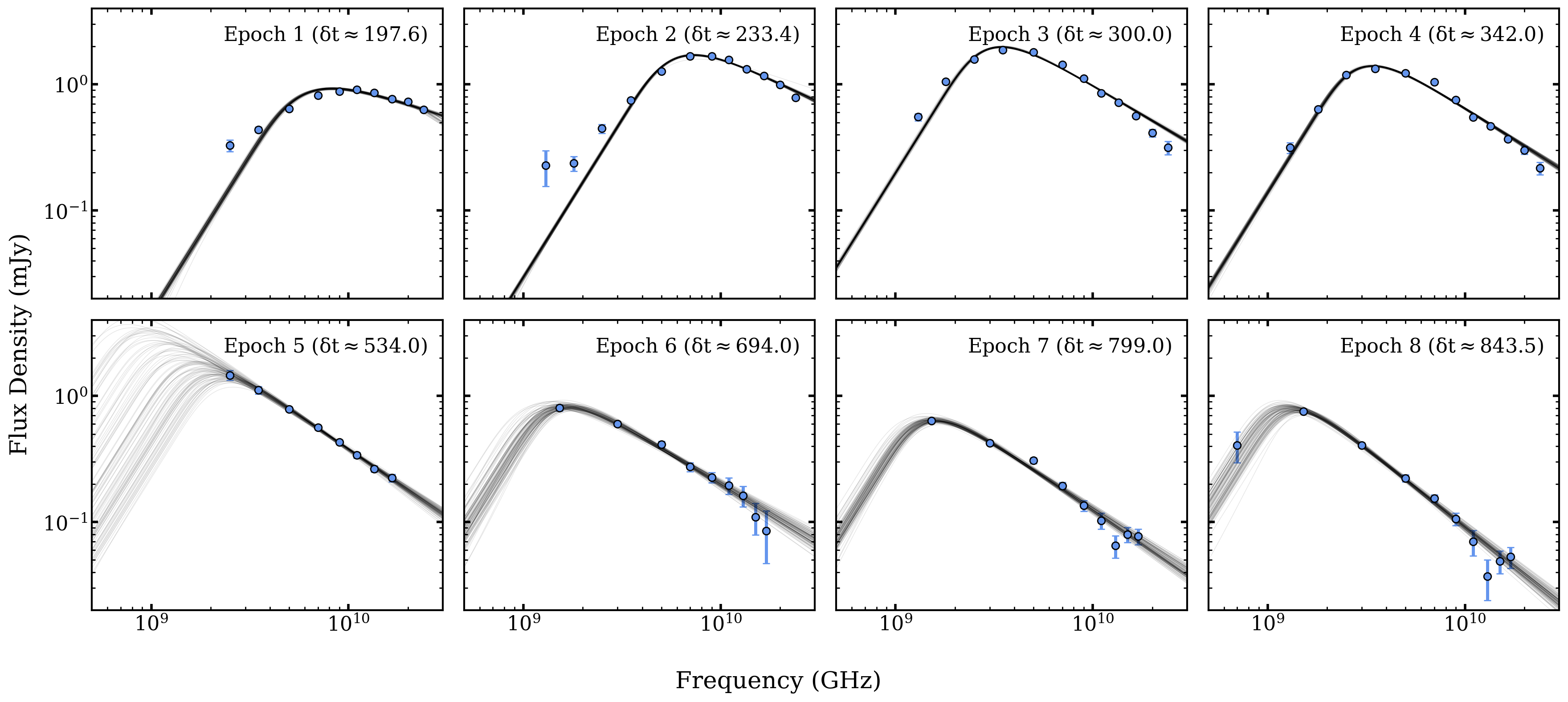}
    \caption{Best fit \mq\ models for the radio SEDs of AT\,2022wtn. The blue points are the observations and the grey lines are the last 1,000 models in the MCMC chain.}
    \label{fig:thermal_electron_bestfit}
\end{figure}

\renewcommand{\thetable}{A2}
\begin{table}[]
    \centering
    \caption{Best fit parameters from the \mq\ model}
    \begin{tabular}{llllll}
    \toprule
    $\delta t$ & p & $\log_{10} \beta \Gamma$ & $\log_{10} n$ & $\log_{10} \epsilon_T$ & $\delta t_{outflow}$ \\
    \midrule
    197.58 & $2.04^{+0.03}_{-0.02}$ & $-0.79^{+0.12}_{-0.10}$ & $3.27^{+0.16}_{-0.25}$ & $-2.33^{+1.10}_{-1.16}$ & $113.18^{+27.79}_{-27.91}$ \\
    233.45 & $2.51^{+0.05}_{-0.05}$ & $-0.79^{+0.17}_{-0.12}$ & $2.87^{+0.23}_{-0.35}$ & $-2.23^{+1.11}_{-1.20}$ & $162.26^{+51.37}_{-53.76}$ \\
    300.00 & $2.86^{+0.03}_{-0.03}$ & $-0.54^{+0.12}_{-0.09}$ & $1.79^{+0.19}_{-0.25}$ & $-2.52^{+0.99}_{-1.00}$ & $216.76^{+54.71}_{-54.97}$ \\
    342.00 & $2.98^{+0.04}_{-0.04}$ & $-0.67^{+0.11}_{-0.08}$ & $2.13^{+0.16}_{-0.22}$ & $-2.45^{+1.09}_{-1.07}$ & $262.37^{+53.97}_{-59.70}$ \\
    534.00 & $3.13^{+0.10}_{-0.09}$ & $-0.55^{+0.26}_{-0.15}$ & $1.33^{+0.57}_{-1.01}$ & $-2.70^{+1.07}_{-0.87}$ & $457.24^{+54.54}_{-61.12}$ \\
    694.00 & $2.90^{+0.13}_{-0.13}$ & $-0.84^{+0.05}_{-0.05}$ & $1.87^{+0.16}_{-0.19}$ & $-2.11^{+1.34}_{-1.30}$ & $608.01^{+57.74}_{-56.54}$ \\
    799.00 & $3.11^{+0.09}_{-0.09}$ & $-0.92^{+0.04}_{-0.04}$ & $2.09^{+0.12}_{-0.14}$ & $-2.15^{+1.28}_{-1.25}$ & $713.17^{+58.02}_{-56.29}$ \\
    841.00 & $3.50^{+0.11}_{-0.10}$ & $-0.79^{+0.05}_{-0.05}$ & $1.80^{+0.18}_{-0.19}$ & $-2.43^{+1.05}_{-1.06}$ & $758.31^{+57.02}_{-58.42}$ \\
    \bottomrule
    \end{tabular}
    \label{tab:thermal-electron-results}
\end{table}

\begin{figure}
    \centering
    \includegraphics[width=0.9\linewidth]{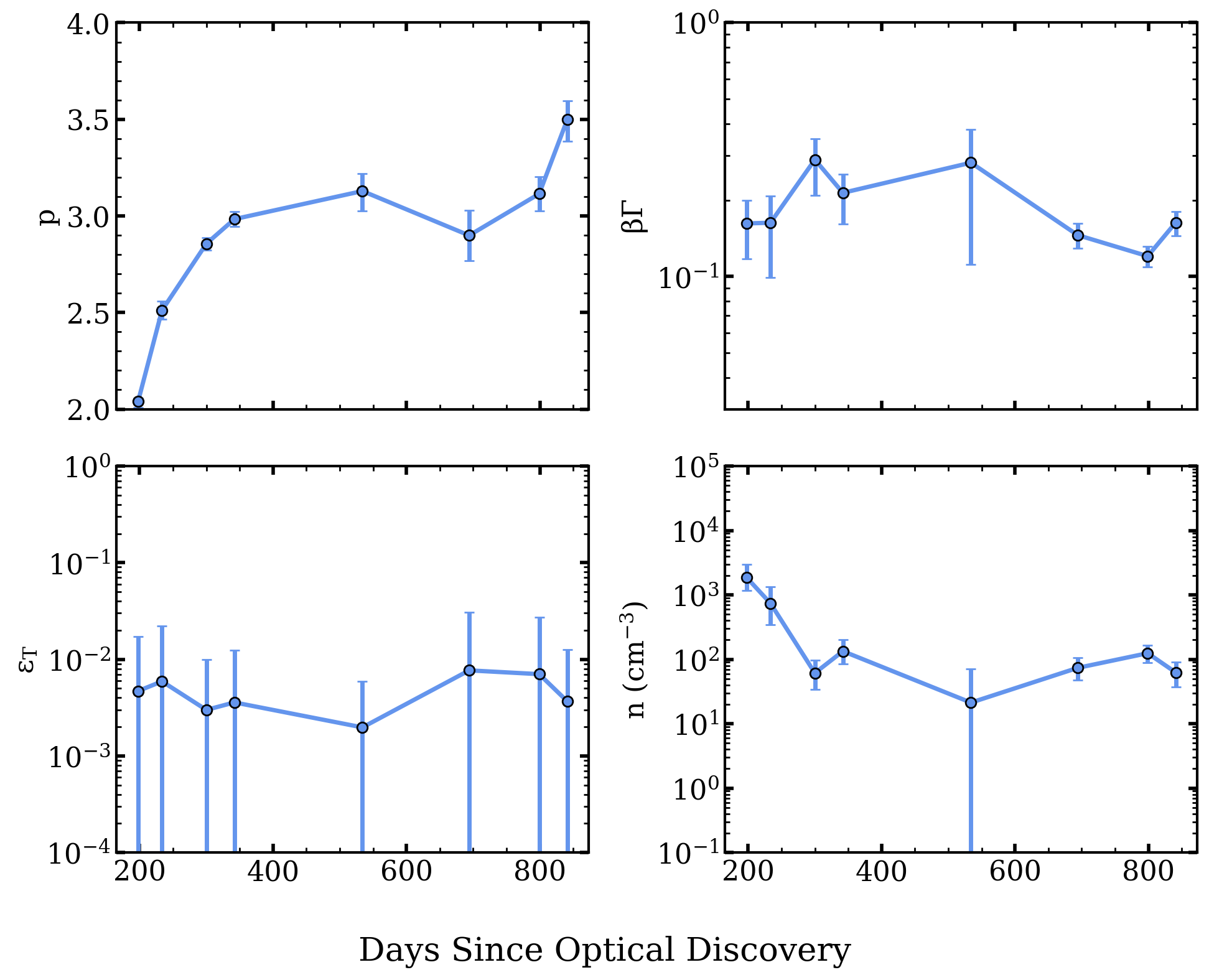}
    \caption{The temporal evolution of the best fit parameters resulting from a fit of the \mq\ model to our data. Points are the median value of the chain and errorbars represent the 68\% uncertainty.}
    \label{fig:thermal-electron-results}
\end{figure}

We fit the radio SEDs independent of each other with the \mq\ model using {\tt emcee} \citep{emcee_main, emcee_ascl}, making an additional assumption of equipartition following \autoref{sec:omodeling}: we fix $\epsilon_e = 0.1$ and solve $\autoref{eq:epsB}$ at each iteration for $\epsilon_B$. For the first four parameters, we use the following uniform priors: $p \in [2,4]$, $\log_{10} \beta\Gamma \in [-4, 0]$, $\log_{10} \left( n / {\rm cm}^{-3} \right) \in [-1, 5]$, $\log_{10}\epsilon_T \in [-4, 0]$. For $\delta t_{\rm outflow}$, we use a uniform prior per epoch, bounded below by the time since first radio detection (169~days), and bounded above by the time since optical discovery, $\delta t_{\rm outflow}\in [\delta t_{\rm obs}-169, \delta t_{\rm obs}]$. SED fits to this model are shown in \autoref{fig:thermal_electron_bestfit}. The median best-fit parameters are given in Table~\ref{tab:thermal-electron-results} and \autoref{fig:thermal-electron-results}.

The inferred values of $\epsilon_T \sim 10^{-2}$ are at least an order of magnitude lower than those of $\epsilon_e$ and $\epsilon_B$ (\autoref{fig:thermal-electron-results}), indicating that impact of thermal electrons to the fits is marginal. We find $p\sim2-3.5$, which is roughly consistent with the mean value $p\approx3.26$ found in \autoref{sec:emodeling}. We find $\beta\Gamma \sim 0.1-0.2$, which is consistent with the values inferred for the spherical outflow case (\autoref{fig:equipartition}). Overall, these results are broadly consistent with the previous analysis in this work. This agreement demonstrates that in the $\epsilon_T \ll 1$ regime the \mq\ model roughly aligns with \citet{matsumoto_generalized_2023}. 

\bibliographystyle{aasjournalv7}
\bibliography{refs}

\end{document}

%% file: authors.tex
\author[orcid=0009-0008-5392-4190,sname='Gavin Farley']{Gavin Farley}
\affiliation{Department of Physics \& Astronomy, University of Utah, Salt Lake City, UT 84112, USA}
\email[show]{gavin.farley@utah.edu}  



\author[orcid=0000-0003-1792-2338,sname='Tanmoy Laskar']{Tanmoy Laskar}
\affiliation{Department of Physics \& Astronomy, University of Utah, Salt Lake City, UT 84112, USA}
\email{tanmoy.laskar@utah.edu}

\author[orcid=0000-0003-4537-3575, gname=Noah, sname=Franz]{Noah Franz}
\email{nfranz@arizona.edu}
\affiliation{Department of Astronomy and Steward Observatory, University of Arizona, 933 North Cherry Avenue, Tucson, AZ 85721-0065, USA}

\author[0000-0003-0528-202X]{Collin T. Christy}
\affiliation{Department of Astronomy and Steward Observatory, University of Arizona, 933 North Cherry Avenue, Tucson, AZ 85721-0065, USA}
\email{collinchristy@arizona.edu}

\author[0009-0007-3919-8439]{Coleman Rohde}
\affiliation{Department of Physics \& Astronomy, University of Utah, Salt Lake City, UT 84112, USA}
\email{u1298655@utah.edu}

\author[0000-0003-3441-8299]{A. J. Goodwin}
\affiliation{International Centre for Radio Astronomy Research – Curtin University, GPO Box U1987, Perth, WA 6845, Australia}
\email{adelle.goodwin@curtin.edu.au}

\author[0000-0002-8297-2473]{Kate D. Alexander}
\affiliation{Department of Astronomy and Steward Observatory, University of Arizona, 933 North Cherry Avenue, Tucson, AZ 85721-0065, USA}
\email{kdalexander@arizona.edu}

\author[0000-0002-9392-9681]{Edo Berger}
\affiliation{Center for Astrophysics \textbar{} Harvard \& Smithsonian, 60 Garden Street, Cambridge, MA 02138-1516, USA}
\email{eberger@cfa.harvard.edu}

\author[0000-0001-7007-6295]{Yvette Cendes}
\affiliation{Department of Physics, University of Oregon, 1371 E 13th Ave, Eugene OR 97403, USA}
\affiliation{Institute for Fundamental Science, University of Oregon, 1371 E 13th Ave, Eugene OR 97403, USA}
\email{yncendes@uoregon.edu}

\author[0000-0002-7706-5668]{Ryan Chornock}
\affiliation{Department of Astronomy, University of California, Berkeley, CA 94720-3411, USA}
\affiliation{Berkeley Center for Multi-messenger Research on Astrophysical Transients and Outreach (Multi-RAPTOR), University of California, Berkeley, CA 94720-3411, USA}
\email{chornock@berkeley.edu}

\author[0000-0003-0307-9984]{Tarraneh Eftekhari}
\affiliation{Center for Interdisciplinary Exploration and Research in Astrophysics (CIERA), Northwestern University, 1800 Sherman Avenue, Evanston, IL 60201, USA }
\email{teftekhari@northwestern.edu}

\author[orcid=0000-0001-7946-1034,sname='Golay',gname='Walter W.']{Walter W. Golay}
\affiliation{Center for Astrophysics \textbar{} Harvard \& Smithsonian, 60 Garden Street, Cambridge, MA 02138-1516, USA}
\email{wgolay@cfa.harvard.edu}

\author[0000-0002-1568-7461]{Wenbin Lu}
\affiliation{Department of Astronomy, University of California, Berkeley, CA 94720-3411, USA}
\email{wenbinlu@berkeley.edu}

\author[0000-0003-4768-7586]{Raffaella Margutti}
\affiliation{Department of Astronomy, University of California, Berkeley, CA 94720-3411, USA}
\affiliation{Department of Physics, University of California, Berkeley, CA 94720-7300, USA}
\affiliation{Berkeley Center for Multi-messenger Research on Astrophysical Transients and Outreach (Multi-RAPTOR), University of California, Berkeley, CA 94720-3411, USA}
\email{rmargutti@berkeley.edu}